\documentclass[12pt,a4paper]{article}

\RequirePackage[l2tabu, orthodox]{nag}

\usepackage{mjheppub}
\usepackage{lipsum}
\usepackage{amsthm,amssymb,amsmath,epic,eepic,float}
\usepackage{rotating,epsfig,indentfirst,array,varioref}
\usepackage{appendix,marginnote,bbm,tikz,pgf,mathtools}
\usepackage{setspace}

\usepackage{tikz}
\usetikzlibrary{decorations.pathreplacing}
\usetikzlibrary{calc}

\usepackage{color}

\let\emptyset\varnothing

\def\ll{\left\lgroup}
\def\rr{\right\rgroup}

\def\leq{\leqslant}
\def\geq{\geqslant}

\newcommand{\ZZ}{\mathbb{Z}}

\newcommand{\slN}{\mathfrak{sl}_N}
\newcommand{\slthree}{\mathfrak{sl}_3}

\newdimen\tableauside\tableauside=1.0ex
\newdimen\tableaurule\tableaurule=0.4pt
\newdimen\tableaustep
\def\phantomhrule#1{\hbox{\vbox to0pt{\hrule height\tableaurule
width#1\vss}}}
\def\phantomvrule#1{\vbox{\hbox to0pt{\vrule width\tableaurule
height#1\hss}}}
\def\sqr{\vbox{%
  \phantomhrule\tableaustep

\hbox{\phantomvrule\tableaustep\kern\tableaustep\phantomvrule\tableaustep}%
  \hbox{\vbox{\phantomhrule\tableauside}\kern-\tableaurule}}}
\def\squares#1{\hbox{\count0=#1\noindent\loop\sqr
  \advance\count0 by-1 \ifnum\count0>0\repeat}}
\def\tableau#1{\vcenter{\offinterlineskip
  \tableaustep=\tableauside\advance\tableaustep by-\tableaurule
  \kern\normallineskip\hbox
    {\kern\normallineskip\vbox
      {\gettableau#1 0 }%
     \kern\normallineskip\kern\tableaurule}%
  \kern\normallineskip\kern\tableaurule}}
\def\gettableau#1 {\ifnum#1=0\let\next=\null\else
  \squares{#1}\let\next=\gettableau\fi\next}

\def\be{\begin{equation}}
\def\ee{\end{equation}}
\def\ba{\begin{array}}
\def\ea{\end{array}}

\newcommand{\< }{{\langle}}
\renewcommand{\>}{{\rangle}}

\restylefloat{figure}

\newcommand{\cB}{\mathcal{B}}
\newcommand{\cC}{\mathcal{C}}

\newcommand{\cF}{\mathcal{F}}

\newcommand{\cH}{\mathcal{H}}

\newcommand{\cL}{\mathcal{L}}
\newcommand{\cM}{\mathcal{M}}
\newcommand{\cN}{\mathcal{N}}
\newcommand{\cO}{\mathcal{O}}

\newcommand{\cT}{\mathcal{T}}

\newcommand{\cW}{\mathcal{W}}

\def\ll{ \left\lgroup}
\def\rr{\right\rgroup}

\newcommand{\product }{\prod }

\begin{document}

\title{\boldmath 
Correlation functions with fusion-channel multiplicity 
in $\cW_3$ Toda field theory
} 

\author{Vladimir Belavin  $^1$,}
\author{Benoit Estienne   $^2$,}
\author{Omar Foda         $^3$,}
\author{Raoul Santachiara $^4$ }
\affiliation{\vspace{2mm} $^1$ 
I E Tamm Department of Theoretical Physics, P N Lebedev Physical Institute,
Leninsky Avenue 53, 119991 Moscow, Russia, 
\newline
Department of Quantum Physics, Institute for Information Transmission Problems,
Bolshoy Karetny per. 19, 127994 Moscow, Russia}
\affiliation{$^2$ LPTHE, CNRS and Universit\'{e} Pierre et Marie Curie, Sorbonne 
Universit\'es, 75252 Paris Cedex 05, France}
\affiliation{$^3$ School of Mathematics and Statistics, University of Melbourne, 
Parkville, Victoria 3010, Australia}
\affiliation{$^4$ LPTMS, CNRS (UMR 8626), Universit\'e Paris-Saclay, 91405 Orsay, 
France}

\emailAdd{
belavin@lpi.ru, 
omar.foda@unimelb.edu.au, 
raoul.santachiara@u-psud.fr
}

\abstract{
Current studies of $\cW_N$ Toda field theory focus on correlation functions 
such that the $\cW_N$ highest-weight representations in the fusion channels 
are multiplicity-free. In this work, we study $\cW_3$ Toda 4-point functions 
with multiplicity in the fusion channel. 
The conformal blocks of these 4-point functions involve matrix elements of 
a fully-degenerate primary field with a highest-weight in the \textit{adjoint} 
representation of $\slthree$, and
a fully-degenerate primary field with a highest-weight in the fundamental      
representation of $\slthree$. 
We show that, when the fusion rules do not involve multiplicities, the matrix 
elements of 
the fully-degenerate adjoint field, between two arbitrary descendant states, 
can be computed explicitly, on equal footing with the matrix elements of the 
semi-degenerate fundamental field.
Using null-state conditions, we obtain a fourth-order Fuchsian differential 
equation for the conformal blocks. Using Okubo theory, we show that, due to 
the presence of multiplicities, this differential equation belongs to a class 
of Fuchsian equations that is different from those that have appeared so far 
in $\cW_N$ theories. We solve this equation, compute its monodromy group, and 
construct the monodromy-invariant correlation functions. This computation shows 
in detail how the ambiguities that are caused by the presence of multiplicities 
are fixed by requiring monodromy-invariance. 
}


\maketitle
\flushbottom

\section{Introduction}
\label{section.01.introduction}

A 2-dimensional conformal field theory is based on the representation theory 
of an infinite-dimensional algebra. In the absence of an extended symmetry, 
the infinite-dimensional algebra is the Virasoro algebra generated by the spin-2 
energy-momentum tensor $\cT(z)$. The most direct generalisations of the Virasoro 
algebra are the $\cW_N$ algebras, $N=2, 3, \cdots$, generated by $(N \! - \! 1)$ 
holomorphic fields of spin-2, $\cdots$, spin-$N$ \cite{fl88}. For a review, see 
\cite{bs95}. 
The main goal in any conformal field theory is to compute the correlation 
functions of the primary fields. One way to achieve this is based on the fact 
that there are null-states in the highest-weight representations that correspond 
to the primary fields \cite{bpz84}. 
This approach allows one to obtain differential equations for the conformal blocks 
which are the building blocks of the correlation function. However, for $\cW_N$ 
models, and particularly for primary fields in higher representations of $\slN$, 
this program contains a number of additional subtleties.

In principle, using operator product expansions, any conformal block can be 
constructed as a series expansion in its holomorphic coordinates. To this end, 
one needs to compute matrix elements of the primary fields between two arbitrary 
descendant states. In $\cW_N$ theories, as opposed to Virasoro theories, 
multiplicities can appear in the tensor products of the irreducible highest-weight 
representations. Consequently, the $\cW_N$ Ward identities are not sufficient to 
express the 3-point functions that involve descendant states in terms of 3-point 
functions that involve primary states only. 
This has, so far, limited the applicability of this approach to correlation 
functions in which all fields, except two, are \textit{semi-degenerate}, with 
(at least) one level-one null-state \cite{fl07c, fl08}. These semi-degenerate 
fields belong to the \textit{fundamental} representation of $\slN$. 
In \cite{kms10}, Kanno \textit{et. al} showed that all 3-point matrix elements that 
contain one semi-degenerate fundamental field can be expressed in terms of 3-point 
functions of primary fields. 

Aside from the above approach, the $\slN$ Coulomb gas approach \cite{fl88, fl07c, 
fl08, furlan2015some} provides a representation of $\cW_N$ correlation functions 
in terms of multi-dimensional integrals, provided that a charge-neutrality 
condition is obeyed. The Coulomb gas correlation functions contain primary fields 
in general $\slN$ representations. However, the evaluation of these multi-dimensional 
integrals is in general possible only in the absence of multiplicities.

To the best of our knowledge, all conformal blocks, currently available in 
the literature \textit{in explicit form}, are such that the multiplicities are not 
present in the fusion channel 
\cite{mironov2010agt, bonelli2012vertices, gl14, alekseev2015wilson}. The purpose 
of this paper is to provide explicit results for correlation functions with such 
multiplicities. 
We focus on $\cW_3$ Toda theory, and consider 4-point correlation functions that 
admit highest-weight representations with multiplicities in the fusion-channel. 
These 4-point functions involve 
a fully-degenerate primary field with a charge-vector in the \textit{adjoint} 
representation 
of $\slthree$, and a fully-degenerate primary field with a charge-vector in the 
fundamental representation of $\slthree$. We show that, when the fusion rules 
are obeyed and do not involve multiplicities, all matrix elements of the 
fully-degenerate adjoint field, between two arbitrary descendant states, can be 
explicitly computed in terms of matrix elements of primary fields. For completeness, 
we provide a parallel discussion of the explicit computation of all matrix elements 
of the semi-degenerate fundamental field, following \cite{kms10}. Given these matrix 
elements, we obtain explicit series expansion for the conformal blocks. 

Next, we show that the conformal blocks formed from the above matrix elements 
satisfy a fourth-order Fuchsian differential equation. We verify that, with the 
appropriate choice of the parameters, the series expansion of the conformal block, 
obtained matrix elements, agrees with the differential equation. 

The remaining part of the paper is devoted to constructing the modular-invariant
correlation function, that corresponds to the previously-computed matrix elements
and conformal blocks, on the basis of the analysis of the differential equation. 
Using Okubo's matrix form \cite{kohno2012global} of the Fuchsian differential 
equations,  
we find that, notwithstanding the presence of an integral difference among the 
characteristic exponents, this equation is free from accessory parameters and 
belongs to type-II in Okubo's classification. Because the fusion rules are obeyed, 
the conformal blocks can be written in terms of Coulomb gas integrals. Using 
the Coulomb gas integral representation, we compute explicitly the monodromy 
group of the differential equation, and show that requiring monodromy invariance  
resolves an ambiguity that arises in the matrix elements when states from 
a highest-weight representation with multiplicity are involved. 

The paper is organised as follows. In Section 
\textbf{\ref{section.02.w3.toda.conformal.field.theory}}, 
we introduce our notation and recall basic facts regarding $\cW_3$ conformal 
field theory.
In \textbf{\ref{section.03.matrix.elements}},
we compute the matrix elements of the fully-degenerate adjoint field between 
two arbitrary descendant states, and discuss the construction of the conformal 
blocks.
In         \textbf{\ref{section.03.fuchsian.differential.equation}}, 
we focus on a specific 4-point correlation function in $\cW_3$ theory, 
and derive a fourth-order Fuchsian differential equation that the conformal 
blocks, that are the building blocks of the correlation function, satisfy.  
In         \textbf{\ref{section.04.coulomb.gas}}, 
we discuss construction of the monodromy-invariant correlation functions using 
the differential equation found in \textbf{\ref{section.03.fuchsian.differential.equation}}.
In         \textbf{\ref{section.06.summary.comments}},
we present our conclusion and discuss open problems.
In appendix \textbf{\ref{appendix.A.Shapovalov}}, 
we discuss the Shapovalov matrix elements that we need in computations.
In          \textbf{\ref{appendix.B.matrix.elements}},
we compute matrix elements.
In          \textbf{\ref{appendix.C.null.states}},
we discuss $\cW_3$ highest-weight modules with null-states at level-2.
In          \textbf{\ref{appendix.D.basis}},
we outline an algorithm to expand any $\cW_3$ state in terms of basis states.

\section{$\cW_3$ conformal field theory}
\label{section.02.w3.toda.conformal.field.theory}

\subsection{$\cW_3$ chiral symmetry algebra}
\label{w3.algebra}
$\cW_3$ is an associative algebra generated by the modes $L_n$ and $W_n$ of the spin-$2$ 
energy-momentum tensor $\cT (z)$ and of the spin-$3$ holomorphic field $\cW(z)$. These 
modes are defined by their action on a field $\Phi(w)$ that corresponds to a state in 
$\cW_3$ theory, 

\begin{eqnarray}
L_n \Phi(w) &=& \frac{1}{2\pi i} \oint_{\cC_w} dz \, (z-w)^{n+1} \, \cT(z) \, \Phi(w),  
\\
W_n \Phi(w) &=& \frac{1}{2\pi i} \oint_{\cC_w} dz \, (z-w)^{n+2} \, \cW(z) \, \Phi(w).
\end{eqnarray}
 
\noindent   
The Virasoro algebra 
\begin{equation}
\left[ L_m, L_n \right]  =  (m-n) L_{m+n} + \frac{c}{12} (m^3-m) \delta_{m+n, 0},
\label{LL.commutator}
\end{equation}
appears as a subalgebra of the $\cW_3$ one. The parametrisation of the $\cW_3$ central 
charge $c$, commonly used in the Toda field theory literature, is  

\begin{equation}
c = 2 + 24 \;Q^2, \quad 
Q = b + \frac{1}{b},
\end{equation} 

\noindent where $Q$ is proportional to the background charge. The full $\cW_3$ algebra 
is given by the following two commutation relations,

\begin{equation}
\left[ L_m, W_n \right] =  (2m-n) W_{m+n},
\label{LW.commutator}
\end{equation}

\begin{multline}
\left[ W_m, W_n \right] = \frac13 (m-n) \Lambda_{m+n} 
\\
+
\ll \frac{22+5c}{48} \rr \ll \frac{m-n}{30}       \rr (2m^2-m n+2 n^2-8) \, L_{m+n}
\\
+
\ll \frac{22+5c}{48} \rr \ll \frac{c}{3 \cdot 5!} \rr (m^2-4) (m^3-m) \, \delta_{m+n, 0}\;,
\label{WW.commutator}
\end{multline}

\noindent where $\Lambda_m$ are the modes of the quasi-primary field 
$\Lambda = :\cT^2: - \frac{3}{10} \, \partial^2 \cT$, 
the colons $: \ :$ stand for normal-ordering, and 

\begin{equation}
\Lambda_m = \sum_{p \leq -2} L_p  L_{m-p}  
          + \sum_{p \geq -1} L_{m-p} L_p  
          - \frac{3}{10} (m+2) (m+3) L_m\;.
\end{equation}

\noindent In \eqref{WW.commutator} we are assuming the following normalisation 
for the $\cW_3$ 2-point function: 

\begin{equation}
\label{norm_w}
\langle \cW (1) \cW (0) \rangle = \eta \equiv \ll \frac{22+5c}{48} \rr\!.
\end{equation}

\paragraph{$\cW_3$ highest-weight modules.}

A $\cW_3$ primary field $\Phi_{h, q}(z)$ corresponds to a state $|h, q \rangle$ 
which is labelled by the eigenvalues, $h$ and $q$, of $L_0$ and $W_0$,

\begin{equation}
L_0 |h,q\rangle = h |h,q\rangle, \quad W_0 |h,q\rangle = q |h,q\rangle.
\end{equation}

\noindent and which is annihilated by all positive modes of $\cT (z)$ and $\cW (z)$,

\begin{equation}
\forall n>0, \quad L_n |h,q\rangle = 0, \quad W_n |h,q\rangle = 0
\end{equation}

\noindent A $\cW_3$ highest-weight representation is spanned by the basis states,

\begin{equation}
\cL_I|h, q \rangle \equiv\; L_{-i_m} \cdots L_{-i_1} 
                            W_{-j_n} \cdots W_{-j_1} |h, q \rangle ,
\label{W3Verma}                            
\end{equation}

\noindent where the sets of positive integers 
$I=\{i_m, \cdots, i_1 ; j_n, \cdots, j_1\}$, are normal-ordered,  

\begin{equation}
i_m \geq \cdots \geq i_1 \geq 1, \quad 
j_n \geq \cdots \geq j_1 \geq 1 .
\end{equation}

\noindent We use 
$\{\emptyset; j_n,\cdots,j_1\}$ and 
$\{i_n,\cdots,i_1; \emptyset\}$ when no $L_i$, and $W_i$ modes act 
on the state, respectively, and $I = \{\emptyset, \emptyset\}$, when 
neither type of modes act on the state, 
$\cL_{\emptyset} \; |h, q \rangle = |h, q\rangle$. In the following 
we will use the notation $\Phi^{(I)_X} =\cL_I \, \Phi_X$ to denote 
a descendant field associated to the primary field 
$\Phi_{X}$, where $X$ indexes the quantum numbers $h,q$. For instance 
$\Phi^{(\{2,1;3,1,1\})_1} =L_{-2}L_{-1}W_{-3}W_{-1}^2 \Phi_{h_1,q_1}$
That any $\cW_3$ highest-weight representation is spanned by the states 
(\ref{W3Verma}), is typically stated without proof in the literature. 
For completeness, we outline a proof in \textbf{\ref{appendix.D.basis}}.

\paragraph{$\cW_3$ Ward identities.} To construct the correlation functions 
of the primary fields we will need the conformal Ward identities. For the 
correlation function with additional insertion of the stress-energy tensor 
$\cT(z)$ one finds

\begin{equation}
\label{T:Ward1}
\left< \cT(z) \prod_{i=1}^{N}\Phi_i(z_i)\right> = 
\sum_{i=1}^N
\ll 
\frac{\Delta_i}{(z-z_i)^2} + \frac{1}{z-z_i} \partial_{z_i}
\rr 
\left< \prod_{i=1}^{N}\Phi_i(z_i)\right>,
\end{equation}

\noindent where $\Phi_i\equiv\Phi_{h_i, q_i}$. From the asymptotic behavior of 
$\cT(z)$, $\underset{z\to \infty}{\lim} \cT(z) \sim 1/z^4$, Equation 
\eqref{T:Ward1} implies the following three identities,

\begin{align}
\label{T: ward2_1}
&\sum_{j=0}^N \partial_{z_j} \left<\prod_{i=1}^N \Phi_i(z_i)\right> =0, \\
&\sum_{j=0}^N 
\ll  z_j \partial_{z_j} + \Delta_j  \rr
\left< \prod_{i=1}^N \Phi_i(z_i) \right> =0, 
\\
\label{T: ward2_3}
& \sum_{j=0}^N \ll z_j^2 \partial_{z_j} +  2 z_j \Delta_j \rr 
\left< 
\prod_{i=1}^N \Phi_i(z_i)
\right>
= 0.
\end{align}

\noindent In analogous way, the Ward identity associated to the conserved current 
$\cW(z)$ takes the form

\begin{equation}
\label{W: ward1}
\left< \cW(z) \prod_{i=1}^N \Phi_i(z_i)\right> =
\sum_{i=1}^N 
\ll 
\frac{q_i}{(z-z_i)^3}+\frac{W^{(i)}_{-1}}{(z-z_i)^2}+\frac{W^{(i)}_{-2}}{z-z_i} 
\rr
\left<\prod_{i=1}^N \Phi_i(z_i)\right>, 
\end{equation}

\noindent where $W^{(i)}_{-n}$ is the mode $W_{-n}$ applied to the field $\Phi_i$. 
From the asymptotic condition $\underset{z\to \infty}{\lim} \cW(z)\to z^{-6}$ one 
obtains additional five Ward identities in $\cW_3$ case, 

\begin{align}
\label{W: ward2_1}
&\sum_{j=0}^N           \, W_{-2}^{(j)}                                     \left<\prod_{i=1}^N \Phi_i(z_i)\right> =0, \\
&\sum_{j=0}^N \ll z_j   \, W_{-2}^{(j)} +         \, W_{-1}^{(j)}             \rr\left<\prod_{i=1}^N \Phi_i(z_i)\right> =0, \\
&\sum_{j=0}^N \ll z_j^2 \, W_{-2}^{(j)} + 2 z_j   \, W_{-1}^{(j)}+   q_j \rr \left<\prod_{i=1}^N \Phi_i(z_i)\right>=0,\\
&\sum_{j=0}^N \ll z_j^3 \, W_{-2}^{(j)} + 3 z_j^2 \, W_{-1}^{(j)}+3 z_j   q_j \rr\left<\prod_{i=1}^N \Phi_i(z_i)\right> =0, \\
\label{W: ward2_5}
&\sum_{j=0}^N \ll z_j^4 W_{-2}^{(j)} + 4 z_j^3 W_{-1}^{(j)}+6 z_j^2 q_j \rr \left<\prod_{i=1}^N \Phi_i(z_i)\right>=0.
\end{align}

\subsection{Highest-weight modules}

The representation theory of the $\cW_3$ algebra  is strictly related to the classical Lie algebra 
$\slthree$. We start this section with some facts about $\slthree$ algebra that are relevant for 
further discussion.

\paragraph{$\slthree$ modules.}
The Lie algebra $\slthree$ is defined by the Cartan matrix $A$, 

\begin{equation}
A = 
\ll 
\begin{array}{cc} 2 & -1 \\ -1 & 2 \end{array} 
\rr
\end{equation}

\noindent the elements of which are the scalar products, $A_{ij} = \langle \vec{e}_i, \vec{e}_j \rangle$, 
of the two simple roots $\vec{e}_1, \vec{e}_2$.

\noindent $A_{11}=A_{22}=2$ and $A_{12}=A_{21}=-1$. The Weyl vector $\vec{\rho}$ 
is half the sum of the positive roots $\{\vec{e}_1, \vec{e}_2, \vec{e}_1+\vec{e}_2\}$, 
that is,
\begin{equation}
\label{vecrho}
\vec{\rho} =   \ll \vec{e}_1+\vec{e}_2 \rr
\end{equation} 
The fundamental weights $\vec{\omega}_1$, $\vec{\omega}_2$ form a basis dual to the simple roots one
\begin{equation}
\langle \vec{\omega}_i, \vec{e}_j \rangle = \delta_{i j}
\end{equation}
The irreducible finite-dimensional representations of $\slthree$ are parametrized by a highest-weight 
on the lattice $\mathbb{N} \vec{\omega}_1 +\mathbb{N} \vec{\omega}_2$ spanned by the fundamental weights. 
In particular, $\vec{\omega}_1$ is the highest-weight of the fundamental representation of $\slthree$, with weights $\vec{h}_i$, $i = 1, 2, 3$,  
\begin{equation}
\label{weight_fund}
\vec{h}_1 =  \vec{\omega}_1, \quad 
\vec{h}_2 = -\vec{\omega}_1 + \vec{\omega}_2, \quad 
\vec{h}_3 = -\vec{\omega}_2,
\end{equation}

\noindent A generic $\slthree$ representation, of highest-weight 
$\vec{\Lambda}$, is indexed by two non-negative integers 
$(\lambda_1, \lambda_2)$ such that

\begin{equation}
\vec{\Lambda} = \lambda_1\; \vec{\omega}_1 + \lambda_2\; \vec{\omega}_2,
\end{equation}

\noindent or equivalently by Young diagram $(1^{\lambda_1},2^{\lambda_2})$, with
$\lambda_k$ columns of length $k$, $(k=1,2)$. The weights of $\vec{\Lambda}$ are 
obtained by filling the cells of the Young diagram with integers $1, 2, 3$, in
non-decreasing order along the rows, and in strictly-increasing order along the 
columns, in all possible manners. Each weight is then obtained by associating 
a cell filled with the integer $i$ to $\vec{h}_i$ and summing over all the cells 
of the Young diagram. The number of the diagrams associated with a given weight 
defines the multiplicity of this weight. For instance, the weights of the fundamental
representation are

\begin{equation}
\label{hi}
\begin{tikzpicture} 
[line width=1.2pt]
\draw (0.00, 0.00) rectangle (0.50, 0.50);
\draw (0.25,-0.05) node[above]{1};
\draw (0.25,-0.75) node[above]{$\vec{\omega}_1$};
\draw (2.00, 0.00) rectangle (2.50, 0.50);
\draw (2.25,-0.05) node[above]{2};
\draw (2.25,-0.75) node[above]{$-\vec{\omega}_1+ \vec{\omega}_2$};
\draw (4.00, 0.00) rectangle (4.50, 0.50);
\draw (4.25,-0.05) node[above]{3};
\draw (4.25,-0.75) node[above]{$-\vec{\omega}_2$};
\end{tikzpicture}
\end{equation}
There is no multiplicity in this case. In the adjoint representation, 
the highest-weight vector is $\vec{\Lambda} = \vec{\omega}_1 +\vec{\omega}_2$,  
and the weights are labelled by Young tableaux as,

\begin{eqnarray}
&& \begin{tikzpicture}
[line width=1.2pt]
\draw (0.0,0.0) rectangle (0.5, 0.5);
\draw (0.0,0.5) rectangle (0.5, 1.0);
\draw (0.5,0.5) rectangle (1.0, 1.0);
\draw (0.25,0.0-0.05) node[above]{2};
\draw (0.25,0.5-0.05) node[above]{1};
\draw (0.75,0.5-0.05) node[above]{1};
\draw (0.25,-0.75) node[above]{$\vec{\omega}_1+\vec{\omega}_2$};
\draw (2.0,0.0) rectangle (2.5, 0.5);
\draw (2.0,0.5) rectangle (2.5, 1.0);
\draw (2.5,0.5) rectangle (3.0, 1.0);
\draw (2.25,0.0-0.05) node[above]{2};
\draw (2.25,0.5-0.05) node[above]{1};
\draw (2.75,0.5-0.05) node[above]{2};
\draw (2.25,-0.75) node[above]{$-\vec{\omega}_1+2\vec{\omega}_2$};
\draw[red] (4.0,0.0) rectangle (4.5, 0.5);
\draw[red] (4.0,0.5) rectangle (4.5, 1.0);
\draw[red] (4.5,0.5) rectangle (5.0, 1.0);
\draw (4.25,0.0-0.05) node[above]{2};
\draw (4.25,0.5-0.05) node[above]{1};
\draw (4.75,0.5-0.05) node[above]{3};
\draw (4.25,0.0-0.75) node[above]{$\vec{0}$};
\draw[red] (6.0,0.00) rectangle (6.5, 0.5);
\draw[red] (6.0,0.50) rectangle (6.5, 1.0);
\draw[red] (6.5,0.50) rectangle (7.0, 1.0);
\draw (6.25,0.0-0.05) node[above]{3};
\draw (6.25,0.5-0.05) node[above]{1};
\draw (6.75,0.5-0.05) node[above]{2};
\draw (6.25,0.0-0.75) node[above]{$\vec{0}$};
\end{tikzpicture} 
\nonumber  
\\
&&\!\!\begin{tikzpicture}
[line width=1.2pt]
\draw (0.0,0.0) rectangle (0.5, 0.5);
\draw (0.0,0.5) rectangle (0.5, 1.0);
\draw (0.5,0.5) rectangle (1.0, 1.0);
\draw (0.25,0.0-0.05) node[above]{3};
\draw (0.25,0.5-0.05) node[above]{1};
\draw (0.75,0.5-0.05) node[above]{1};
\draw (0.25,-0.75) node[above]{$2\vec{\omega}_1-\vec{\omega}_2$};
\draw (2.0,0.0) rectangle (2.5, 0.5);
\draw (2.0,0.5) rectangle (2.5, 1.0);
\draw (2.5,0.5) rectangle (3.0, 1.0);
\draw (2.25,0.0-0.05) node[above]{3};
\draw (2.25,0.5-0.05) node[above]{1};
\draw (2.75,0.5-0.05) node[above]{3};
\draw (2.25,-0.75) node[above]{$\vec{\omega}_1-2\vec{\omega}_2$};
\draw (4.0,0.0) rectangle (4.5, 0.5);
\draw (4.0,0.5) rectangle (4.5, 1.0);
\draw (4.5,0.5) rectangle (5.0, 1.0);
\draw (4.25,0.0-0.05) node[above]{3};
\draw (4.25,0.5-0.05) node[above]{2};
\draw (4.75,0.5-0.05) node[above]{2};
\draw (4.25,-0.75) node[above]{$-2\vec{\omega}_1 + \vec{\omega}_2$};
\draw (6.0,0.0) rectangle (6.5, 0.5);
\draw (6.0,0.5) rectangle (6.5, 1.0);
\draw (6.5,0.5) rectangle (7.0, 1.0);
\draw (6.25,0.0-0.05) node[above]{3};
\draw (6.25,0.5-0.05) node[above]{2};
\draw (6.75,0.5-0.05) node[above]{3};
\draw (6.25,-0.75) node[above]{$- \vec{\omega}_1 - \vec{\omega}_2$};
\end{tikzpicture} 
\end{eqnarray}

\noindent We note that the weight $\vec{0}$ has multiplicity two.

\paragraph{$\cW_3$ primary operators.}

Each primary field $\Phi_{h, q}(z)$ can be labelled 
by a vector $\vec{\alpha}$ in the space spanned by the fundamental $\slthree$ weights,

\begin{equation}
\vec{\alpha} = \alpha_1 \; \vec{\omega}_1 + \alpha_2 \; \vec{\omega}_2 
\end{equation}

\noindent  The vector charge 
$\vec{\alpha}$ can be written in terms of the vector charge $\vec{P}$ as

\begin{equation}
\vec{\alpha} = \vec{P} + Q \vec{\rho}
\end{equation} 
where $\vec{\rho}$ has been defined in \eqref{vecrho}. 
\noindent Introducing the parameters 

\begin{equation}
x_i =  \vec{P} \cdot \vec{h}_i,  \quad i = 1, 2, 3, 
\end{equation}

\noindent where $\vec{h}_i$ is defined in (\ref{weight_fund}), the quantum numbers, 
$h$ and $q$, are 

\begin{align}
	h & = Q^2 + x_1 x_2 + x_1 x_3 + x_2 x_3   = Q^2 - \frac{1}{2} \ll x_1^2 + x_2^2 + x_3^2 \rr, 
	\\
 q & = i \; x_1 x_2 x_3,
\end{align}

\noindent The above expressions are invariant under the $\slthree$ Weyl group action. 
This group is composed by six elements $\hat{s}(\vec{P})$ which act on $\vec{P}=(P_1,P_2)$ 
in the following way,

\begin{eqnarray}
 (P_1,P_2) &\to & (-P_1, P_1+P_2) \to (P_2, -P_1 - P_2) \to  \nonumber \\ &\to&
  (-P_2, -P_1)\to (P_1 + P_2, -P_2) \to  (-P_1 - P_2, P_1)
\end{eqnarray} 

\noindent Henceforth, we will use equivalently the pair $(h, q)$ or the symbol $\vec{\alpha}$ 
to indicate a primary field. As the charge $\vec{\alpha}$ and its Weyl reflections $\vec{\alpha}-Q\vec{\rho}$ correspond to the same pair  $(h, q)$, the notation 

\begin{equation}\label{refl_sym}
\Phi_{h, q}(z) \equiv\Phi_{\vec{\alpha}}(z)\equiv \Phi_{Q\vec{\rho}+ 
\hat{s}(\vec{\alpha}-Q\vec{\rho})}(z) 
\end{equation}

\noindent indicate the same primary field. We will use also the notation:

\begin{equation}
\label{dual}
\Phi^{*}_{\vec{\alpha}}(z)\equiv \Phi_{h, -q}(z) \equiv \Phi_{2Q \vec{\rho}- \vec{\alpha}}(z)
\end{equation}

\noindent to indicate the dual field $\Phi^{*}_{\vec{\alpha}}$ which satisfies

\begin{equation}
\langle \Phi_{\vec{\alpha}}|\Phi_{\vec{\alpha}}\rangle = \lim_{z\to \infty} z^{2 h}\langle  
\Phi^*_{\vec{\alpha}} (z) \Phi_{\vec{\alpha}} (0) \rangle =1.
\end{equation}

\noindent The above relation fixes the normalization of the primary operators of the theory. 
Note that the vector $2 Q \vec{\rho}-\vec{\alpha}$ is associated to $\vec{P}=(-P_1,-P_2)$. 

The charge parametrization of the primary operator is reminiscent of their Coulomb 
gas representation, introduced  in Section \ref{CGcomp}. In particular, when one 
represents a primary operator with a vertex operator in a Coulomb gas theory, the 
identification  (\ref{refl_sym}) implies non-trivial normalisations factors, {\it cf}  
(\ref{refamp.1}).

\paragraph{Structure constants of the diagonal $\cW_3$ theory.}
The symmetry group of the $\cW_3$ theory is the tensor product $\cW_3\otimes \bar{\cW_3}$ 
of holomorphic and anti-holomorphic algebras. The spectrum of the diagonal theory, which 
is built from the $\cW_3\otimes \bar{\cW_3}$ representations, is composed by the primary 
fields 

\begin{equation}
\Phi^{\text{phys}}_{\vec{\alpha}}(z, \bar{z}) \equiv
\Phi_{\vec{\alpha}}(z)\Phi_{\vec{\alpha}}(\bar{z}).
\end{equation} 

\noindent The operator product expansion, OPE, of two primary fields takes the form

\begin{equation}
\Phi^{\text{phys}}_{\vec{\alpha}_M}(z,\bar{z}) \Phi^{\text{phys}}_{\vec{\alpha}_R}(0) = 
\sum_L |z|^{-2h_R-2h_M+2h_L } C_{\vec{\alpha}_M,\vec{\alpha}_R}^{\vec{\alpha}_L} 
\left[\Phi^{\text{phys}}_{\vec{\alpha}_L}(z,\bar{z})\right], 
\end{equation}

\noindent where $\left[\Phi^{\text{phys}}_{\vec{\alpha}_L}(z, \bar{z})\right]$ denotes 
the contribution of the primary field and all its descendants and the summation in the
above equation goes over all the possible primaries $\Phi^{\text{phys}}_{L}$ (that can
constitute a continuous set). The structure constants 
$C_{\vec{\alpha}_M, \vec{\alpha}_R}^{\vec{\alpha}_L}$ can be defined by the 3-point
function,
 
\begin{equation}
\label{threepoint}
C_{\vec{\alpha}_M, \vec{\alpha}_R}^{\vec{\alpha}_L} = 
\left< \Phi^{\text{phys}}_L \Big| \Phi^{\text{phys}}_M (1)\Phi^{\text{phys}}_R (0)\right>.
\end{equation}

\paragraph*{Fully-degenerate representations.}
Each $\cW_3$ fully-degenerate representation is associated with a primary 
field $\Phi_{\vec{\alpha}}(z)$, with a vector-valued charge $\vec{\alpha}$, 

\begin{equation}
\vec{\alpha}_{r_1 r_2 s_1 s_2} = 
b
\ll
(1 - r_1)\; \vec{\omega}_1 +
(1 - r_2)\; \vec{\omega}_2 
\rr 
+ 
\frac{1}{b}
\ll
(1 - s_1)\; \vec{\omega}_1 +
(1 - s_2)\; \vec{\omega}_2 
\rr,
\end{equation}

\noindent where $r_1, r_2, s_1,s_2$ are positive integers. We denote this primary field 
by $\Phi_{r_1 r_2 s_1 s_2} (z)$. The highest-weight representation $\mathcal{V}_{r_1 r_2 s_1 s_2}$ 
associated to the field $\Phi_{r_1 r_2 s_1 s_2}$ exhibits two independent null-states 
with charges 
$\vec{\alpha}_{r_1, r_2, s_1, s_2} + b r_1 \vec{e}_1$ and 
$\vec{\alpha}_{r_1, r_2, s_1, s_2} + b r_2 \vec{e}_2$.
The fusion products of $\mathcal{V}_{r_1 r_2 s_1 s_2}$ with a general $\cW_3$ irreducible 
module $\mathcal{V}_{\vec{\alpha}}$ takes the form
\begin{equation}
\label{w3: fuprod}
\mathcal{V}_{r_1 r_2 s_1 s_2} \times \mathcal{V}_{\vec{\alpha}} =  
\sum_{\vec{h}_r, \vec{h}_s} \mathcal{V}_{\alpha - b \vec{h}_r-b^{-1} \vec{h}_s},
\end{equation}

\noindent where $h_r$ and $h_s$ are the weights of the $\slthree$ representation with 
highest-weight \\$(r_1-1)\; \vec{\omega}_1 +
( r_2-2)\; \vec{\omega}_2$ and $(s_1-1)\; \vec{\omega}_1 +
( s_2-2)\; \vec{\omega}_2$ respectively.

\subsection{The Coulomb gas approach} 
\label{CGrepre}

We now briefly describe the Coulomb gas approach to $\cW_3$ Toda theory, which will 
be relevant in constructing conformal blocks in Section \textbf{\ref{section.04.coulomb.gas}}.

In the Coulomb gas approach, a primary field $\Phi^{\text{phys}}_{\vec{\alpha}}$ is represented 
by the exponential field 
$V^{\text{phys}}_{\vec{\alpha}} (z,\bar{z}) = :e^{\vec{\alpha} \cdot \vec{\phi} (z,\bar{z})}:$, where $\vec{\phi}$ is a two-component free boson field. In particular, we have the identification

\begin{equation}
\label{refamp.1}
\Phi^{\text{phys}}_{\vec{\alpha}} = N^{-1}_{\vec{\alpha}} V_{\vec{\alpha}}(z, \bar{z}),
\end{equation}

\noindent where the normalisation constant $N^{-1}_{\vec{\alpha}}$ has been computed in \cite{fat01}. 
For a given $\vec{\alpha}$, several vertex operators can be used, since we have the identification 

\begin{equation}
\label{refamp.2}
V_{\vec{\alpha}}  = R_{s(\vec{\alpha})}  V_{s(\vec{\alpha})},
\end{equation}

\noindent where $R_{s(\vec{\alpha})}$ is the \textit{reflection amplitude} in Toda theory \cite{fat01}.
The free boson correlation function is,

\begin{equation} 
\left< \prod_{i=1}^{N} V^{\text{phys}}_{\vec{\alpha}_i}(z_i,\bar{z}_i)\right>_{\!\!\!\text{free boson}} 
= \delta_{ \ll \sum_{i=1}^N \vec{\alpha}_i, \ 2 Q \vec{\rho} \rr}
\prod_{1\leq i<j\leq N}|z_i-z_j|^{-4\vec{\alpha}_i \cdot \vec{\alpha}_j},
\end{equation}

\noindent where $\delta_{a,b}$ is the Kronecker delta and the background charge $2Q\vec{\rho}$ 
is at infinity. This method provides the following integral representation for the $\cW_3$  
correlation functions on the plane,

\be
\label{CGcomp}
\left\langle \product_{i=1}^{N}\Phi^{\text{phys}}_{\vec{\alpha}_i}(z_i,\bar{z})\right\rangle\propto
\left\langle 
\mathcal{Q}_1^{n_1} (b)
\mathcal{Q}_1^{m_1} (b^{-1})
\mathcal{Q}_2^{n_2} (b)
\mathcal{Q}_2^{m_2} (b^{-1})
\product_{i=1}^N
V^{\text{phys}}_{\vec{\alpha}_i} (z_i, \bar{z}_i)\right\rangle_{\!\!\!\text{free boson}}\;,
\ee

\noindent where $\mathcal{Q}_{1,2}(b)$ are screening operators

\be
\mathcal{Q}_k(b)=\int e^{b \vec{e}_k \cdot \vec{\phi}(z, \bar{z})} d^2 z\;,
\ee

\noindent if the neutrality condition

\begin{equation}
\label{neutrality}
\sum_{i=1}^{N} \vec{\alpha}_i +  \ll n_1 b+ m_1 b^{-1} \rr \vec{e}_1 +  
                                 \ll n_2 b+ m_2 b^{-1} \rr \vec{e}_2 = 
                                 2Q \vec{\rho}\;,
\end{equation}

\noindent is fulfilled for some natural $n_k,m_k$, $k=1,2$.

A Coulomb gas approach can be used to determine directly the holomorphic components 
of the correlation function, that is, the conformal block. We refer the reader to 
Chapter \textbf{8} of \cite{dot98} for a detailed description of this approach. 
In this case, one formally considers the holomorphic part of the exponential field
$V^{\text{phys}}_{\vec{\alpha}}(z,\bar{z}) \to V_{\vec{\alpha}}(z)$, and associates 
the holomorphic primary field to it,  $\Phi_{\vec{\alpha}_i}=V_{\vec{\alpha}}(z)$.
The exponential free boson correlator reads,

\begin{equation}
\label{CGhol}
\left< \prod_{i=1}^{N} V_{\vec{\alpha}_i}(z_i)\right>_{\!\!\!\text{free boson}} = 
\delta_{\ll \sum_{i=1}^N \vec{\alpha}_i, \ 2 Q \vec{\rho} \rr} 
\prod_{1\leq i<j\leq N}(z_i-z_j)^{-2\vec{\alpha}_i \cdot
\vec{\alpha}_j}.
\end{equation}

\noindent In the case the neutrality condition \eqref{neutrality} is satisfied, one 
obtains the representation for the conformal block in terms of closed contour integrals,

\begin{multline}
\left\langle \product_{i=1}^{N}\Phi_{\vec{\alpha}_i}(z_i)\right\rangle \propto
\\
\left\langle \prod_{i=1,2} \ll  \oint du  V_{b \, \vec{e}_i} (u)  \rr^{n_i} 
\ll \oint du  V_{b^{-1} \, \vec{e}_i} (u) \rr^{m_i} \prod_{i=1}^N
V_{\vec{\alpha}_i}(z_i) 
\right\rangle_{\!\!\!\text{free boson}}
\end{multline}

\noindent The different choices of the contours are in correspondence with the different 
fusion channels, as shown in our specific case below.  For more details on Coulomb gas 
approach in $\slthree$ Toda theory, we refer the reader to \cite{fl88,fl07,fl08}.

\section{Conformal blocks and matrix elements}
\label{section.03.matrix.elements}

\noindent In the following, we define a general $\cW_3$ 4-point conformal block.

\subsection{Power series expansion of the conformal block}

Using invariance under global conformal transformations 
(\ref{T: ward2_1}--\ref{T: ward2_3}), we fix $z_R=0, z_2=1$, and $z_L=\infty$, 
and consider the 4-point conformal block $\cB_M \ll L, 2, 1, R \rr (z)$ as 
a function of the holomorphic coordinate $z_1 = z$. The fusion channel is 
labelled $M$ in the \textit{comb diagram},

\begin{equation}
\label{combdiag}
\begin{tikzpicture}
[line width=1.2pt]
\draw (-4.5,0)node[left]{$\cB_M \ll L, 2, 1, R \rr (z)$};
\draw (-4.5,0)node[right]{$\equiv$};
\draw (-4,0)node[right]{$\langle \Phi_{L}|\Phi_2 (1) \Phi_1 (z) | \Phi_R \rangle$};
\draw (0,0) node[right]{$\equiv$};
\draw (2,0)--(3,0);
\draw (3,0)--(3,1);
\draw (3,0)--(5,0);
\draw (5,0)--(5,1);
\draw (5,0)--(6,0);
\draw (2,0) node[left]{$\Phi^*_L (\infty)$};
\draw (3,1) node[above]{$\Phi_2 (1)$};
\draw (5,1) node[above]{$\Phi_1 (z)$};
\draw (6,0) node[right]{$\Phi_R (0)$};
\draw (4,0) node[below]{$M$};
\end{tikzpicture}
\end{equation}

\noindent The most general $\cW_3$ 4-point conformal block depends on ten parameters, 
since each primary field, as well as the highest-weight representation that flows in 
the fusion channel, is specified by two quantum numbers, $h$ and $q$. 
In the presence of multiplicities, the ten parameters are not sufficient to determine 
the conformal block, as we explain later on our specific example (see the discussion 
below \eqref{X2}). Using standard techniques, one can express the conformal block as 
a series expansion in $z$,

\begin{multline}
z^{h_1 + h_R - h_M} \cB_M \ll L, 2, 1, R \rr (z)
=
\\
1 + 
\sum_{i=1}^\infty z^i 
\sum_{
\substack{
K_M, K_M' 
\\ 
|K_M| = |K'_M| = i
}
} 
\ll H^{-1}\rr_{K_M, K'_M} \Gamma_{\emptyset_L, \emptyset_2, K_M}
 \Gamma^{\prime}_{K'_M, \emptyset_1, \emptyset_R},
\label{cb.series.expansion}
\end{multline}

\noindent where the main ingredients are the elements of the Shapovalov matrix of inner 
products, $H$, and the matrix elements of the primary fields between two arbitrary 
descendant states, $\Gamma_{I_L,J_M,K_R}$ and $\Gamma^{\prime}_{I_L, J_M, K_R}$. 

\subsection{The Shapovalov matrix of inner products}
The Shapovalov matrix $H$, whose $ij$-element $H_{ij}$ is the scalar product of the 
states
$|I \rangle_{h, q} = \cL_I |h, q \rangle $ and 
$|J \rangle_{h, q} = \cL_J |h, q \rangle $, defined in terms of the basis 
(\ref{W3Verma}),

\begin{equation}
H_{IJ} = \, _{h, q} \langle I |J \rangle_{h, q}\;.
\end{equation}

\noindent $H$ is a block-diagonal matrix, $\text{diag} \ll H_0, H_1, H_2, \cdots \rr$, 
where the elements of the $i$-th block, $H_i$, are the scalar products of the level-$i$
descendants. These elements can be computed using the commutation relations
(\ref{WW.commutator}). By definition, $H_0 = 1$, and the explicit forms of $H_1$ and $H_2$
are given in Appendix \textbf{\ref{appendix.A.Shapovalov}}. 

\subsection{The matrix elements}
\label{kanno.algorithm}
The second ingredient in \eqref{cb.series.expansion} is the matrix elements. Using 
the notation of \cite{kms10}, we are interested in the matrix elements of general
descendant fields

\begin{eqnarray}
\label{gamma-gammap} 
\Gamma_{I_L, J_M, K_R} 
 &=& 
\frac{\langle 
(\Phi^{*}_{L})^{(I)_L}| 
\, 
\Phi^{(J)_M}_{M}(1) 
\, 
\Phi^{(K)_R}_{R}(0)
\rangle}{\langle 
(\Phi^{*}_{L})| 
\, 
\Phi_{M}(1) 
\, 
\Phi_{R}(0)
\rangle}\;,
\\
\Gamma^{\prime}_{I_L, J_M, K_R}  &=& 
\frac{\langle 
\Phi^{(I)_L}_{L}| 
\, 
\Phi^{(J)_M}_{M}(1) 
\, 
\Phi^{(K)_R}_{R}(0)
\rangle}{\langle 
\Phi_{L}| 
\, 
\Phi_{M}(1) 
\, 
\Phi_{R}(0)
\rangle}  \;,
\end{eqnarray}

\noindent where the descendants fields $\Phi^{(I)_X}_X =\cL_I \, \Phi_X$, $X = L, M, R$, were defined in (\ref{W3Verma}). The matrix elements $\Gamma$ and $\Gamma^{\prime}$ are related by changing the sign of $q_L$.
 
To compute the 4-point conformal block explicitly, we need to express each of 
these matrix elements in terms of matrix elements of three primary fields. This 
can be achieved using Virasoro and $\cW_3$ Ward identities in a systematic way, 
as explained in \cite{fl07c,bw92,bpt92,bw93, wat94, kms10} or in the work of Kanno \textit{et al.} \cite{kms10}, where an algorithm to
compute the matrix elements of descendant fields in terms of the matrix elements 
of the corresponding primaries, was outlined. We review below this procedure

Let us start with the Virasoro Ward identities. These lead 
to three recursion relations for the matrix elements. For $n \in \ZZ > 0$, 
one of these recursion relations is

\begin{multline}
\< L_{-n} \, \Phi^{(I)_L}_L \, |\, \Phi^{(J)_M}_M \, |\, \Phi^{(K)_R}_R \> =
\\
\< L_0 \Phi^{(I)_L}_L  \, |\,  \Phi^{(J)_M}_M \, |\,  \Phi^{(K)_R}_R \> +
\<     \Phi^{(I)_L}_L  \, |\,  \Phi^{(J)_M}_M \, |\, 
\ll L_n - L_0 \rr              \Phi^{(K)_R}_R\>                         +
\\
\label{Vir.recursion.L}  
\<     \Phi^{(I)_L}_L \, |\, 
\ll n L_0 + \sum_{i=1}^n \frac{(n+1)!}{(i+1)! (n-i)!} L_i \rr 
       \Phi^{(J)_M}_M  \, |\,  \Phi^{(K)_R}_R \rangle,
\end{multline}

\noindent The point of the recursion relation \eqref{Vir.recursion.L} is that 
the Virasoro creation operator $L_{-n}$, $n>0$, that acts on $\Phi^{(I)_L}_L$ 
on the left hand side of (\ref{Vir.recursion.L}), is replaced by $L_0$, and 
annihilation operators $L_i$, $i>0$, that act on $\Phi^{(J)_M}_M$ and
$\Phi^{(K)_R}_R$ on the right hand side. 
Since the level of the initial descendant $L_{-n} \, \Phi^{(I)_L}_L$, on the 
left hand side is finite, the final descendants on the right hand side can 
be expanded in terms of basis states, in finitely-many steps, as explained 
in \textbf{ \ref{appendix.D.basis}}. 

The result of the application of (\ref{Vir.recursion.L}) is that the initial 
matrix element on the left hand side is expanded in terms of new matrix elements 
on the right hand side. The new  matrix elements are such that each of them 
contains one descendant that is \textit{closer}  to its primary state, while 
the other two are the same.
Thus each of the matrix elements on the right hand side are closer to matrix 
elements of primaries. Applying \eqref{Vir.recursion.L} repeatedly, we can 
compute the matrix element of any descendant on the left hand side in terms of 
the matrix element of the corresponding primaries. Most importantly, we can do 
that in finitely many steps. In each step, we peel a Virasoro creation operator 
off a descendant, and replace it with $L_0$ or an annihilation operator, and 
thereby bring the matrix element closer to that of three primaries.
The two remaining Virasoro recursion relations of Kanno \textit{et al.} are 
analogous to (\ref{Vir.recursion.L}), but involve peeling a Virasoro 
creation operator $L_{-n}$, $n>0$, off $\Phi^{(J)_M}_M$ and off $\Phi^{(K)_R}_R$, 
respectively. They can be found in \cite{kms10}.

Next, we turn to the $\cW_3$ Ward identities. These also lead to three recursion 
relations for the matrix elements. One of these is

\begin{multline}
\< W_{-n} \, \Phi^{(I)_L}_L \, |\, \Phi^{(J)_M}_M \, |\, \Phi^{(K)_R}_R \> =
    \\
\< W_0 \Phi^{(I)_L}_L \, |\,  \Phi^{(J)_M}_M \, |\,  \Phi^{(K)_R}_R \> +
\< \Phi^{(I)_L}_L \, |\,  \Phi^{(J)_M}_M \, |\, \ll W_n - W_0 \rr \Phi^{(K)_R}_R \> +
\\
\< \Phi^{(I)_L}_L \, |\, 
\ll 
n W_{-1} +
\frac12 n (n+3) W_0 + 
\sum_{i=1}^n \frac{(n+2)!}{(i+2)! \, (n-i)!} \, W_i 
\rr 
\Phi^{(J)_M}_M \, |\, \Phi^{(K)_R}_R \>,
\label{W.recursion.L}
\end{multline}

\noindent and two more $\cW_3$ recursion relations that are analogous to
(\ref{W.recursion.L}), but involve peeling a $\cW_3$ creation operator
$W_{-n}$, $n>0$, off $\Phi^{(J)_M}_M$ and $\Phi^{(K)_R}_R$, respectively 
\cite{kms10}. The result of the algorithm of Kanno \textit{et al.} in 
the $\cW_3$ case, is not as simple as in the Virasoro case. In the $\cW_3$ 
case, one can start from any matrix element of three arbitrary $\cW_3$ 
descendant states, $\Phi^{(I)_L}_L$, $\Phi^{(J)_M}_M$, and $\Phi^{(K)_R}_R$, 
and re-write it in terms of the matrix element

\begin{equation}
\langle 
\Phi_{L} 
\, | \, 
\ll W_{-1}^p \Phi_{M}
\rr 
\, | \,
\Phi_R
\rangle, 
\quad 
p = 0, 1, 2, \cdots
\label{result.1}
\end{equation}

\noindent the evaluation of which, for arbitrary $\Phi_M$, is non-trivial, 
and generally not explicitly known. 

\paragraph{Matrix elements of the semi-degenerate fundamental field.}

A primary field $\Phi_{\vec{\alpha}}(z)$  in the fundamental representation 
of $\slthree$  has a charge $\vec{\alpha}$ that is collinear to $\vec{\omega}_1$,

\begin{equation}
\vec{\alpha} = a\; \vec{\omega}_1, 
\end{equation}

\noindent and quantum numbers

\begin{equation}
\label{haqa}
h_a = 
\frac{a}{3 b}
\ll 3+ 3 b^2- a b  \rr, \quad 
q_a = \frac{i a}{27 b^2} \ll 3+3b^2 -a b \rr  \ll 3+3 b^2-2 ab \rr.
\end{equation}

\noindent We consider first the case where the fusion field $\Phi_M$ in the above 
matrix elements is associated to the fundamental representation, 

\begin{equation}
\label{Mwy}
\Phi_M = \Phi_{a \vec{\omega}_1}.
\end{equation} 

\noindent It can be shown that the corresponding highest-weight representation has 
a null-state at level-one, so that in general $\Phi_{a\vec{\omega}_1}$ is 
semi-degenerate. Imposing the vanishing of the null-state, one obtains the identity 

\begin{equation}
W_{-1} \Phi_{a \vec{\omega}_1} = \frac{3 q_a}{2h_a} L_{-1} \Phi_{a \vec{\omega}_1}.
\label{Level.1.identity}
\end{equation}

In \cite{kms10}, Kanno \textit{et al.} use this identity to compute the matrix 
elements \eqref{result.1} of $ \Phi_{M}$, defined as in \eqref{Mwy}. The procedure 
is as follows. We replace the string $W_{-1}^p$ by the string $W_{-1}^{p-1} \, L_{-1}$, 
expand the operator $W_{-1}^{p-1} \, L_{-1}$ in terms of the basis (\ref{W3Verma}), 
then apply the algorithm to peel all Virasoro creation operators, $L_{-n}, n > 0$, 
and all $\cW_3$ creation operators, $W_{-n}, n > 1$, that act on $ \Phi_{M}$ in 
the resulting matrix elements, and re-express these as diagonal and annihilation 
operators on $\Phi_{L}$ or $\Phi_{R}$. The action of the diagonal operators 
on the primary fields is known, and the action of the annihilation operators is 
zero by definition. Repeating this procedure, Kanno \textit{et al.} express 
any matrix element of a primary field with a highest-weight in the fundamental 
of $\slthree$, in terms of the matrix element of the corresponding primaries. 
For instance,

\begin{equation}
\langle 
\Phi_{L}
\, | \,
\ll 
W_{-1} \Phi_{a \, \vec{\omega}_1} 
\rr
\, | \,
\Phi_{R}
\rangle =
\frac32 \frac{q_a}{h_a} 
\ll -h_a+h_L-h_R \rr  
\langle 
\Phi_L
\, | \,
\Phi_{a \, \vec{\omega}_1}
\, | \,
\Phi_R
\rangle,
\end{equation}

\noindent and 

\begin{multline}
\langle 
\Phi_L
\, | \,
\ll W_{-1}^2 \Phi_{a \, \vec{\omega}_1} \rr 
\, | \,
\Phi_R
\ \rangle  = 
\\
\frac{3 q_a}{2 h_a} 
\ll
\frac{3 q_a}{2 h_a} (-h_a+h_L-h_R)(-h_a+h_L-h_R-3)-q_L-q_R-q_a
\rr
\\
\langle 
\Phi_{L}
\, | \,
\Phi_{a \, \vec{\omega}_1} 
\, | \,
\Phi_R
\rangle.
\end{multline}

\noindent For the purposes of the present work, we will use this algorithm 
in the case $a = -b$. This corresponds to identifying the fusion field 
$\Phi_M$ with the fully-degenerate fundamental field 
$\Phi_{-b \vec{\omega}_1} = \Phi_{2111}$.
In addition to the level-1 null-state \eqref{Level.1.identity}, this field 
also obeys level-2 and level-3 null-state conditions, 

\begin{equation}
     W_{-2} \Phi_{-b \vec{\omega}_1}  =\frac{1}{\sqrt{\eta}} 
\ll \frac{12 q             }{h (5 h + 1)} L_{-1}^2 - 
    \frac{ 6 q (h+1)}{h (5 h + 1)} L_{-2} \rr   \Phi_{-b \vec{\omega}_1} ,
\label{Level.2.identity}
\end{equation}

\noindent and

\begin{multline}
\label{Level.3.identity}
W_{-3}\Phi_{-b \vec{\omega}_1}  =
\frac{1}{\sqrt{\eta}}\ll\frac{16 q}{ h_M (h+ 1) (5 h +1)} L_{-1}^3
\right.
\\
\left.
- \, \frac{12 q}{h (5 h + 1)} L_{-1} L_{-2}-\frac{3q(h-3)}{2 h (5 h + 1)}
L_{-3} \rr\Phi_{-b \vec{\omega}_1},
\end{multline}

\noindent where $h$ and $q$ are given respectively by $h_a$ and $q_a$ in 
the \eqref{haqa} with $a=-b$,

\begin{equation}
\label{hq2111}
h= h_{2111} = 
\frac13 \ll - 3 - 4 b^2 \rr,
\quad 
q = q_{2111} = 
- \frac{i}{27 b} \ll 3+4b^2 \rr \ll 3 + 5 b^2 \rr
\end{equation}

\noindent and $\eta$ has been defined in \eqref{norm_w}.

\paragraph{Matrix elements of the fully-degenerate adjoint field.}
We are interested in computing the matrix elements, when the primary field $\Phi_{M}$, 
that flows in the fusion channel, is the fully-degenerate primary field $\Phi_{2211}$
\footnote{\,
The result for the primary $\Phi_{1122}$ is obtained from 
the result for $\Phi_{2211}$ by replacing $b \to \frac{1}{b}$.
},

\begin{equation}
\Phi_{M}=\Phi_{2211},
\end{equation}

\noindent with quantum numbers,

\begin{equation}
\label{hq2211}
h=h_{2211}=-2-3 b^2, \quad q=q_{2211}= 0 .
\end{equation}

\noindent The $\Phi_{2211}$ highest-weight representation has two null-states at 
level-two. Considering the states at level-two, we obtain, see Appendix 
\textbf{\ref{appendix.C.null.states}}, the null-state conditions,

\begin{equation}
L_{-1} W_{-1} \Phi_{2211}=\frac12 (h+1)W_{-2} \Phi_{2211},
\label{deg2211}
\end{equation}

\noindent and

\begin{equation}
W_{-1}^2 \Phi_{2211}=\frac{1}{\eta}
\ll
-\frac{2(h+2)(h+1) h}{(h-3)(5 h+1)} L_{-2} + 
\frac{3(h+1)(h-1)}{2(h-3)(5 h+1)}L_{-1}^2 
\rr
\Phi_{2211}.
\label{level.2.identity}
\end{equation}

\noindent From (\ref{level.2.identity}), the action of $W_{-1}^2$ on $\Phi_{2211}$ 
can be re-written in terms of the action of Virasoro operators. In analogy with 
the fundamental case, using the algorithm of Kanno \textit{et al.}, together with 
the identity (\ref{level.2.identity}), one can compute all the matrix elements of
$\Phi_{2211}$. More specifically, using the algorithm, one can re-write any matrix 
element $\langle \Phi_{L}^{(I)} \, | \, \Phi_{2211} \, | \,  \Phi_{R}^{(K)} \rangle$
in terms of

\begin{equation}
\langle 
\Phi_{L}^{} 
\, | \, 
\ll 
W_{-1}^p \Phi_{2211} 
\rr 
\, | \,
\Phi_{R}^{}
\rangle, 
\quad 
p = 0, 1, 2, \cdots,
\label{result.2}
\end{equation}

\noindent Next, one applies (\ref{level.2.identity}) to reduce the degree $p$ to 
the degree $(p-2)$. In the case $p=2$, one obtains

\begin{multline}
\langle \Phi_L| \ll W_{-1}^2 \Phi_{2211}(1) \rr |\Phi_R\rangle =
\\
\frac{1}{\eta} 
\ll 
\frac{
3 (h + 1) (h - 1) (-h + h_L - h_R) (-h + h_L -h_R - 1)
}{
2 (h - 3) (5 h + 1)
}
\right.
\\
\left.
-\frac{ 2 h (h + 1)(h + 2)(h -h_L + 2 h_R) 
}{ 
(h - 3)(5 h + 1) 
}
\rr
\langle 
\Phi_{L}
\, | \, 
\Phi_{2211}
\, | \,
\Phi_R
\rangle.
\end{multline}

\noindent  When $p>2$, 
one needs to apply (\ref{level.2.identity}) finitely-many times. Following every 
application, one obtains two matrix elements of states that are not in basis form.
These states need to be expanded in terms of basis states (\ref{W3Verma}). Following 
that, one applies the algorithm of Kanno \textit{et al.} to each of these states. 
Since $p$ is initially finite, one obtains, in finitely-many steps, a matrix element 
of three primaries, or a linear combination of the latter and the matrix element
(\ref{result.2}), but now with $p=1$. Using the relation (\ref{deg2211}) together with the identities (\ref{T: ward2_1})-(\ref{W: ward2_5}), the latter can be explicitly evaluated as,

\begin{equation}
\label{W1}
\langle 
\Phi_L
\, | \,
\ll 
W_{-1} \Phi_{2211}
\rr
\, | \, 
\Phi_R
\rangle
= 
\frac{(q_R-q_L)(h +1)}{2(h_R-h_L)}
\langle 
\Phi_L        
\, | \,
\Phi_{2211}
\, | \, 
\Phi_R
\rangle.
\end{equation}
When $\Phi_L= \Phi_R$ the  matrix element
\begin{equation}
\left< \Phi_{L}| \ll W_{-1} \Phi_{2211}\rr |\Phi_{L}\right>
\end{equation}

\noindent is not defined. This matrix element is related to the channel
$\vec{\alpha}_L = \vec{\alpha}_R + \vec{0}$ and therefore to the two-fold 
degenerate $\vec{0}$ weight in the adjoint representation. The same phenomenon 
happens in the classical $\slthree$ representation theory where the $SU(3)$ 
Clebsch-Gordan coefficients are not uniquely defined when multiplicities are 
present (see for instance \cite{alex2010numerical}).

\section{A fourth-order Fuchsian differential equation}
\label{section.03.fuchsian.differential.equation}

We consider the 4-point conformal block

\begin{equation}\label{Block}
\begin{tikzpicture}
[line width=1.2pt]
\draw (0,0)node[left]{$\cB_M (z)$};
\draw (0,0)node[right]{$\equiv$};
\draw (2,0)--(3,0);
\draw (3,0)--(3,1);
\draw (3,0)--(5,0);
\draw (5,0)--(5,1);
\draw (5,0)--(6,0);
\draw (2,0) node[left ]{$\Phi^{*}_L      (\infty)$};
\draw (3,1) node[above]{$\Phi_{2211} (1     )$};
\draw (5,1) node[above]{$\Phi_{2111} (z     )$};
\draw (6,0) node[right]{$\Phi_R      (0     )$\;,};
\draw (4,0) node[below]{$M$};
\end{tikzpicture}
\end{equation}

\noindent where, for the moment, $\Phi_L$ and $\Phi_R$ are general fields. 
Using the null-state conditions associated with $\Phi_{2211}(1)$, together 
with the null-state conditions associated with $\Phi_{2111}(z)$, we derive 
a fourth-order Fuchsian differential equation for $\cB_M (z)$. Since this 
is a standard procedure (see, for example, \cite{fl08, epss11, bfs15}), it 
suffices to outline the derivation. For sake of clarity, it is convenient 
to re-introduce the explicit dependence of the conformal block on general 
positions $z_i$, $i = 1, 2, 3, 4$ of the external operators,

\begin{equation}
\cB_{M}(\{z_i\}) = 
z_{41}^{-2 h_1} 
z_{42}^{h_1+h_3-h_2-h_4} 
z_{43}^{h_1+h_2-h_3-h_4}
z_{32}^{h_4-h_1-h_2-h_3} 
\cB_M \ll \frac{z_{12} z_{34} }{z_{14} z_{32}} \rr,
\end{equation} 

\noindent where $z_{ij} = z_i - z_j$. The operators 
$\Phi_{2111}, \Phi_{2211}, \Phi_{R}$ 
and $\Phi_L$ are placed at positions $z_1, z_2, z_3$ and $z_4$ respectively. 
The conformal block \eqref{Block} is obtained in the limit 
$z_1\to z$, $z_2 \to 0$, $z_3\to 1$ and $z_4\to \infty$. According to these 
conventions, in the next formula, we use the labelling, 

\begin{eqnarray}
h_1 &=& h_{2111},\quad  q_1 = q_{2111}, \quad h_3 
                            = h_{2211}, \quad q_3 
                            = q_{2211}=0 
\\
h_2 &=& h_R, 
\quad q_2 = q_R, \quad h_4 = h_L, \quad q_4 = - q_L,
\end{eqnarray}

\noindent see \eqref{hq2111} and \eqref{hq2211}. 

\paragraph{The adjoint and fundamental null-state conditions.}

First, we analyze the consequences of the null-state condition \eqref{Level.3.identity}. 
The RHS in \eqref{Level.3.identity} contains only Virasoro generators and does not 
represent any difficulty. Using \eqref{T:Ward1} it can be easily transformed into 
the differential operator acting on $\cB_M \ll \{z_i\} \rr$,

\begin{multline}
\label{RHS}
\text{RHS}  \Rightarrow 
\frac{1}{\sqrt{\eta}} 
\ll \frac{16 q_1}{ h_1 (h_1 + 1) (5 h_1 +1)} \partial_{z_1}^3 
\right.
\\
\left.
- \, \frac{12 q_1}{h_1 (5 h_1 + 1)}  \partial_{z_1} \sum_{k\neq 1 }
\ll
\frac{ h_k}{(z_1-z_k)^2}+
\frac{\partial_{z_k}}{(z_1-z_k)  }
\rr 
\right.
\\
\left.
+ \, \frac{3q_1(h_1-3)}{2 h_1 (5 h_1 + 1)}
\sum_{k \neq 1 }
\ll
\frac{ 2h_k}{(z_1-z_k)^3}+
\frac{\partial_{z_k}}{(z_2-z_k)^2  }
\rr  
\rr 
\cB_M \ll \{ z_i \} 
\rr.
\end{multline}

\noindent For the LHS in \eqref{Level.3.identity} we can use \eqref{W: ward1} to get

\begin{equation}
\label{LHS}
\text{LHS}  \Rightarrow   \sum_{k\neq 1} 
\ll
\frac{q_k        }{(z_1-z_k)^3}+
\frac{W_{-1}^{(k)}}{(z_1-z_k)^2}+ 
\frac{W_{-2}^{(k)}}{(z_1-z_k)  }
\rr\cB_M \ll \{z_i\} \rr.
\end{equation}
 
\noindent Now we have to take into account the $\cW_3$  identities 
(\ref{W: ward2_1}--\ref{W: ward2_5}). These five relations mix nine unknown functions,
$W_{-1}^{(i)} \cB_M \ll \{z_i\} \rr$, $i=1, \cdots, 4$, $W_{-2}^{(i)} \cB_M \ll \{z_i\} \rr$,
$i=1,\cdots,4$ and $\cB_M \ll \{z_i\} \rr$.
The fact that we have at position $z_1$ the field $\Phi_{-b\vec{\omega}_1}$ provides
three more equations, the level-1 and level-2 conditions \eqref{Level.1.identity}, 
and \eqref{Level.2.identity},

\begin{equation}
W_{-1}^{(1)}\cB_M \ll \{z_i\} \rr = \frac{3 q_1}{2 h_1} \partial_{z_1}\cB_M \ll \{z_i\} \rr,
\label{WWard.1}
\end{equation}

\begin{multline}
W_{-2}^{(1)}
\cB_M \ll \{z_i\} \rr= \!\!
\\
\ll \frac{12 q_1             }{h_1 (5 h_1 + 1)} 
\partial_{z_1}^2 -
\frac{ 6 q_1 (h_1+1)}{h_1 (5 h_1 + 1)} \sum_{ k \neq 1 } \!
\ll
\frac{ h_k}{(z_1-z_k)^2} +
\frac{\partial_{z_k}}{(z_1 - z_k)}
\rr    
\!\rr\!\!\cB_M \ll \{z_i\} \rr,
\label{WWard.2}
\end{multline}

\noindent and the level-$3$ equation \eqref{Level.3.identity} which can be expressed
by equating \eqref{RHS} and \eqref{LHS}. Finally, the last equation that we need is
provided by the null-state condition \eqref{deg2211},

\begin{equation}
\partial_{z_3} W_{-1}^{(3)}\cB_M \ll \{z_i\} \rr =
\frac12 (h_{3}+1)W_{-2}^{(3)} \cB_M \ll \{z_i\} \rr.
\label{consq-deg2211}
\end{equation}

Using the previous relations, we obtained  a partial differential equation for
$\cB_M ( \{z_i\} )$. By fixing projective invariance with help of 
(\ref{T: ward2_1}--\ref{T: ward2_3}), this can be transformed in a fourth-order 
Fuchsian equation for $\cB_M(z)$. The explicit form of the differential equation 
is given in \eqref{BF} and in \eqref{DE}.

\paragraph{Fusion rules and local exponents.}

Before giving the explicit form of the fourth-order Fuchsian equation, we first discuss
the general properties of its solutions. Of course, these solutions have to correspond 
to the \textit{\lq admissible\rq\,} fusion, corresponding to the case when $\vec{\alpha}_L$ 
and $\vec{\alpha}_R$ are related through 

\begin{align}
\vec{\alpha}_L + b \, \vec{\beta}_1 + b  \, \vec{\beta}_2 = \vec{\alpha}_R,
\end{align}

\noindent where $\vec{ \beta}_1$ is a weight in the fundamental representation, with 
highest-weight  $    \vec{\omega}_1                     $, and 
                $\vec{ \beta}_2$ is a weight in the adjoint representation, with 
highest-weight  $\ll \vec{\omega}_1 + \vec{\omega}_2 \rr$. 
We expect the case $\vec{\beta}_2 = \vec{0}$, which corresponds to the multiplicity-2 
weight in the adjoint representation to require special attention, so we focus on it. 
Without loss of generality, we choose $\vec{\beta}_1 = \vec{\omega}_1$, the two other 
cases being related it to it by the action of the Weyl group. We end up with, 

\begin{equation}
\label{alar}
\vec{\alpha}_L + b \, \vec{\omega}_1 =  \vec{\alpha}_R.
\end{equation}

\noindent Defining 

\begin{equation}
\vec{\alpha}_R = a_{R_1} \vec{\omega}_1 + a_{R_2} \vec{\omega}_2,
\end{equation}

\noindent the conformal block $\cB_M (z)$ is a function of three parameters, $a_{R_1}$,
$a_{R_2}$ and $b$. The dependence on $M=1, 2, 3, 4$, indexes the four solutions of the
fourth-order Fuchsian differential equation. The possible channels are,

\begin{eqnarray}
\label{ch1}
\text{Channel 1:        }\quad 
\vec{\alpha}_M &=& \vec{\alpha}_R+ b \, \ll \vec{\omega}_1 - \vec{\omega}_2 \rr \\
\label{ch2}
\text{Channel 2:        }\quad 
\vec{\alpha}_M &=& \vec{\alpha}_R+ b \,     \vec{\omega}_2     \\
\label{ch34}
\text{Channel 3 \textit{and} 
      channel 4:        }\quad
\vec{\alpha}_M &=& \vec{\alpha}_R- b \,     \vec{\omega}_1
\end{eqnarray}

\noindent Channels 3 and 4 correspond to the fusion

\begin{equation}
	\Phi_{\vec{\alpha}_L                       } \Phi_{2211} = 
	\Phi_{\vec{\alpha}_R  - b \, \vec{\omega}_1} \Phi_{2211} \to 
	\Phi_{\vec{\alpha}_R  - b \, \vec{\omega}_1},
\end{equation}

\noindent which reflects the fact that, in the adjoint representation, with highest 
weight $\ll \vec{\omega}_1+\vec{\omega}_2 \rr$, there is a weight 
$\vec{\Lambda} = \vec{0}$ with multiplicity 2. This is consistent with the local 
exponents of the Fuchsian differential equation, given below, with singularities 
at $0, 1$ and $\infty$. In Riemann-symbol notation
\footnote{\,
We refer the reader to \cite{yoshida1987fuchsian} for an exhaustive overview of 
Fuchsian systems.},
the local exponents are

\begin{equation}
\label{Riem_not}
\begin{Bmatrix}  
0& & 1 & & \infty  \\ 
\alpha_1   & &  \beta_1   & & \gamma_1   \\
\alpha_2   & &  \beta_2   & & \gamma_2   \\
\alpha_3   & &  \beta_3   & & \gamma_3   \\
\alpha_3+1 & &  \beta_3+1 & & \gamma_3+1 
\end{Bmatrix}
\end{equation}

\noindent with

\begin{eqnarray}
\label{alphas}
\alpha_1 &=&-\frac{  a_{R_1}}{3} b    +\frac{  a_{R_2}}{3} b +  b^2 + 1, 
\quad
\alpha_2 =2 -\frac{  a_{R_1}}{3} b    -\frac{2 a_{R_2}}{3} b +2 b^2,  
\quad   
\alpha_3 = \frac{2 a_{R_1}}{3} b    +\frac{  a_{R_2}}{3} b,  
\nonumber \\
\label{betas}
\beta_1 &=& 1 +   b^2, 
\quad
\beta_2 = 2 + 3 b^2,
\quad
\beta_3 =   -   b^2,\nonumber \\
\label{gammas}
\gamma_1 &=& -2+\frac{  a_{R_1}}{3} b +\frac{2 a_{R_2}}{3} b -3 b^2,\,\, 
\gamma_2 = -1+\frac{a_{R_1}}{3} b -\frac{  a_{R_2}}{3} b -2 b^2,\,\,
\gamma_3 =   -\frac{2 a_{R_1}}{3} b -\frac{  a_{R_2}}{3} b.  \nonumber \\
&&
\end{eqnarray} 

\noindent One can verify that: 

\begin{equation}
\sum_{i=1}^4 \ll \alpha_i + \beta_i + \gamma_i \rr = 6,
\end{equation}

\noindent where $\alpha_4 = \alpha_3+1$, $\beta_4 = \beta_3+1$ and 
$\gamma_4 = \gamma_3+1$. This is the Fuchsian relation for a fourth-order Fuchsian
differential equation with $2+1$ singularities. We see that the multiplicity in 
the representation with highest-weight $\ll \vec{\omega}_1 + \vec{\omega}_2 \rr$ 
reflects in the degeneracy of characteristic exponents, two of which differ by 
an integer. This means that we have a 2-dimensional space spanned by two solutions 
with the same local exponents. In other words, we have a family of solutions (conformal blocks) of 
the fourth-order differential equation whose expansion at the first-order is not 
fixed.  

\paragraph{Explicit form of the fourth-order differential equation.}
We found that the differential equation obeyed by the conformal block \eqref{Block} 
can be given in a simple form  in terms of the function $\cF(z)$, defined as,

\begin{equation}
\label{BF}
\cB_M (z) = z^{\alpha_3}(z-1)^{\beta_3} \cF(z),
\end{equation}

\noindent  For the function  $\cF(z)$ we get

\begin{multline}
z^2 (z-1)^2 \cF''''(z) + z (z-1) \ll b_1 z + c_1 \rr \cF'''(z)
\\
+ \ll a_2 z^2 + b_2 z + c_2 \rr \cF''(z)
+ \ll           b_3 z + c_3 \rr \cF' (z)
+                       c_4     \cF  (z) = 0,
\label{DE}
\end{multline}
where
\begin{align}
c_1& = -2 - 2 a_{R_1} b - a_{R_2} b + 3 b^2,\\
b_1 &= 4 + 2 a_{R_1} b + a_{R_2} b - 9 b^2,\\
c_2 &= b (a_{R_1} + a_{R_2} - 2 b )(1 + a_{R_1} b - b^2)  ,
    \end{align}
   \begin{align}
b_2 &= b (-6 a_{R_1} - 4 a_{R_2} + 17 b - 2 a_{R_1}^2 b - 2 a_{R_1} a_{R_2} b + 
     14 a_{R_1} b^2 + 6 a_{R_2} b^2 - 21 b^3),
     \\
a_2 & = 
b (5 a_{R_1} + 3 a_{R_2} - 19 b + a_{R_1}^2 b + a_{R_1} a_{R_2} b - 11 a_{R_1} b^2 - 
     5 a_{R_2} b^2 + 27 b^3),
\\
c_3 & = b (1 - b^2) (2 a_{R_1} - 3 b - 2 a_{R_1}^2 b - 2 a_{R_1} a_{R_2} b + 
     10 a_{R_1} b^2 + 4 a_{R_2} b^2 - 11 b^3),
\\
b_3 &= b (1 - b^2) (-2 a_{R_1} + 8 b + 2 a_{R_1}^2 b + 2 a_{R_1} a_{R_2} b - 
     16 a_{R_1} b^2 - 7 a_{R_2} b^2 + 31 b^3),
\\
c_4 &= (-1 + b) b^2 (1 + b) (1 - a_{R_1} b + 3 b^2) (2 - a_{R_1} b - a_{R_2} b + 
     4 b^2).
\end{align}

\noindent

\subsection{Okubo theory: classification of Fuchsian systems }

Our goal here is to characterise, in a clear way, the fourth-order Fuchsian equation 
\eqref{DE}. By comparing this equation to the one obtained in \cite{fl07c} for the 
4-point conformal block with a semi-degenerate field, we want to clarify the consequences
of replacing a field in the fundamental representation with a field in the adjoint one.
In \cite{okubo1987group, okubo1971connection} Okubo developed a theory to study the 
global properties of Fuchsian systems. He showed first that any Fuchsian equation can 
be put in a form of a Fuchsian system called the Okubo normal form. This form is 
particularly convenient for the definition of systems that are \textit{\lq accessory-free\rq\,}, 
which are the systems that are uniquely determined, up to trivial transformations, 
by the positions of the singularities and by the differences between the exponents 
associated to each singularity. We refer the reader to \cite{katz1996rigid}, or to 
Chapter \textbf{20} of \cite{inceordinary} for a precise definition of the accessory
parameters. For accessory-free systems, there is an algorithm to determine the
corresponding monodromy group. An introduction to the global properties of Fuchsian 
systems, and to Okubo theory, can be found in \cite{kohno2012global}.

Consider a function $F(z)$ that obeys an order-$n$ Fuchsian equation with $p$ singular
points $z=\mu_i$, $i=1, \cdots, p$. The Okubo form of an order-$n$ Fuchsian equation 
takes the form,

\begin{equation}
\label{okubo.form}
\ll z \mathbb{I}- B \rr \frac{d X(z)}{d z} = A X(z)
\end{equation}

\noindent where $X(z)$ is an $n$-component vector, the components of which are functions
related to $F$ and its derivatives. $\mathbb{I}$ is the $\ll n \! \times \! n \rr$ 
identity matrix, $B$ is a diagonal matrix with $p$ repeated elements, $\mu_i$, 
$i=1, \cdots, p$, 

\begin{equation}
B = \text{diag} 
\ll 
\overbrace{\mu_1, \cdots, \mu_1}^{m_1}, 
\overbrace{\mu_2, \cdots, \mu_2}^{m_2},
\cdots, 
\overbrace{\mu_p, \cdots, \mu_p}^{m_p}
\rr,
\quad
m_1 + m_2 + \cdots + m_p = m, 
\end{equation}

\noindent and $A$ is  $\ll n \! \times \! n \rr$ matrix with constant entries. 
We assume that $A$ is diagonalisable,

\begin{equation}
A \sim \text{diag}
\ll
\overbrace{\nu_1, \cdots, \nu_1}^{n_1},
\overbrace{\nu_2, \cdots, \nu_2}^{n_2}, 
\cdots, 
\overbrace{\nu_q, \cdots, \nu_q}^{n_q}
\rr
\label{Adiag}
\end{equation}

\noindent The eigenvalues of the matrix $A$ correspond to the local exponents 
$\gamma_i$ associated to the singularities at infinity. Corresponding to the 
partition in $B$, the matrix $A$ can be decomposed into 
$\mu_1, \mu_2, \cdots,\mu_p$ matrix blocks $A_{i j}$, $1\leq i,j\leq p$ of size 
$m_i \times m_j$,

\begin{equation}
A = 
\ll
\begin{array}{cccc}
 A_{11} & A_{12}& \cdots & A_{1p} \\
 A_{21} & A_{22}& \cdots & A_{2p}\\
 \vdots & \ddots &\ddots & \vdots\\
 A_{p1}&\cdots &\cdots & A_{pp}
\end{array}
\rr,
\end{equation} 

\noindent We assume further that the diagonal blocks $A_{i i}$, $i=1,\cdots,p$ 
are diagonalisable,

\begin{equation}
A_{ii} \sim \text{diag}
\ll
\lambda^{(i)}_1,\cdots,\lambda^{(i)}_{m_i}
\rr
\quad i = 1,\cdots,p.
\label{Aidiag}
\end{equation}

\noindent The eigenvalues $\lambda^{(i)}_1$ are the local exponents associated 
to the singular points $z=\mu_i$. On the assumptions \eqref{Adiag} and 
\eqref{Aidiag} \cite{haraoka1994canonical}, an irreducible Fuchsian system
\eqref{okubo.form} is accessory-free if \cite{haraoka1994canonical},

\begin{equation}
\cN \equiv n^2 - n + 2 -\sum_{i=1}^p m_i^2 - \sum_{j=1}^q n_j^2 = 0,
\end{equation}

\noindent In Okubo theory, the Fuchsian systems are then characterized by 
the set of integers $m_i$ and $n_i$ describing the multiplicities of the eigenvalues 
of the matrices $B$ and $A$. Consider an order-$n$ Fuchsian equation. There are eight 
classes of systems which are known to be accessory-free,

\begin{itemize}
\item I:  $m_1=m-1, m_2=1$ and $n_1=n_2=\cdots=n_{n}=1$ 

I$^{*}$: $m_1=m_2=\cdots=m_n=1$ and $n_1=n-1, n_{2}=1$ 

\item $(n= 2 l, l\geq 2)$ 

II: $m_1=m_2= l$ and $n_1=l, n_2= l-1, n_3=1$

II$^{*}$: $m_1=l, m_2= l-1, m_3=1$ and $n_1=n_2= l$ 

\item   $(n= 2 l+1, l\geq 2)$ 

III: $m_1= l+1, m_2= l$ and $n_1=l, n_2= l, n_3=1$

III$^{*}$: $m_1=l, m_2= l, m_3=1$ and $n_1= l+1, n_2= l$

\item $(n=6)$

IV: $m_1=4, m_2=2$ and $n_1=n_2=n_3=2$

IV$^{*}$: $m_1=m_2=m_3=2$ and $n_1=4, n_2=2$ 
\end{itemize}

\noindent Class-I corresponds to the generalized hypergeometric equations. 
These are the equations obtained in \cite{fl07c}, or more recently in 
\cite{alekseev2015wilson}. 
The differential equations obeyed by $n+2$-point $\cW_{N}$ conformal blocks 
involving $n_1$ fundamental and $n_2$ anti-fundamental fully-degenerate fields, 
with $n_1 + n_2 = n$, are related to the Calogero-Sutherland Hamiltonian \cite{es09, epss11}.

\subsection*{Okubo form of the fourth-order differential equation}
The differential equation (\ref{DE}) is a Fuchsian differential equation with 
coefficients $a_i, b_i$ and $c_i$ that are functions of $a_{R_1}, a_{R_2}$ and 
$b$. In our case, we have

\begin{equation}
B=\text{diag} \ll 1, 1, 0, 0 \rr,
\quad
\text{and}
\quad
A=
\ll
\begin{array}{cc}
 A_{11} & A_{12}  \\
 A_{21} & A_{22}
\end{array}
\rr,
\end{equation}

\noindent where

\begin{align}
A_{11}=
\ll
\begin{array}{ccc}
  0            & \phantom{11} & 1       \\
 -2 \ll 1+2 b^2 \rr^2 & \phantom{11} & 6 b^2+3 \\
\end{array}
\rr\;,
\quad 
A_{12}=
\ll
\begin{array}{cc}
0 &  0 \\
1 &  0
\end{array}
\rr,
\end{align}

\begin{eqnarray}
\ll A_{21} \rr_{11} 
& = &
(b + b^3) 
\ll
5  (2 a_{R_1} + a_{R_2}) b^2 + 2 (2 a_{R_1} + a_{R_2}) - 4 b^3 - 2b
\rr, 
\nonumber 
\\
\ll A_{21} \rr_{12} & = &
- (b + b^3) \ll 4 a_{R_1} + 2 a_{R_2} - b \rr, 
\nonumber 
\\
\ll A_{21} \rr_{21} & = &
- (b + b^3) 
\ll
a_{R_1}^2 (6b + 13 b^3) + (a_{R_2} - 2 b) 
\ll 
a_{R_2} (2 + 5 b^2) - 2 (b + 2 b^3) 
\rr
b
\rr 
\nonumber 
\\
& & - (b + b^3)
\ll
a_{R_1} \ll 2 + a_{R_2} (6b + 13 b^3) - 3 (b^2 + 7 b^4) \rr
\rr,  
\nonumber 
\\
\ll A_{21} \rr_{22} & = & 
(b + b^3) 
\ll
4 a_{R_1}^2 b + a_{R_1} (4 a_{R_2} b-6 b^2 + 4) + (2 a_{R_2} - b) (a_{R_2} - 2b) b
\rr,
\end{eqnarray}

\begin{align}
A_{22}=
\ll
\begin{array}{ccc}
 0 &  \phantom{11} & 1 
\\
-(a_{R_1} + a_{R_2} - 2 b) 
\ll 1 + (a_{R_1} - b) b \rr b & \phantom{11} & -1 - (2 a_{R_1} + a_{R_2}) b + 3 b^2
\end{array}
\rr,
\end{align}

The matrix $A$ can be diagonalized,

\begin{equation}
A \sim \text{diag}  
\ll 
2 - a_{R_1} b - a_{R_2}\; b + 4 b^2,1 - a_{R_1} b + 3 b^2, b^2, -1 + b^2
\rr
\end{equation}

One can check that these eigenvalues, consistently with \eqref{BF}, correspond to 

\begin{equation}
-\gamma_i-\alpha_2-\beta_3, \quad i=1,2,3
\end{equation}

\noindent and to 

\begin{equation}
-\gamma_3-1-\alpha_2-\beta_3
\end{equation}

\noindent and are therefore associated to the singularity $\gamma_i$ at infinity. 
Two of the eigenvalues differ by an integer. As we show below, we do not have 
logarithmic solutions, which is consistent with the fact that we are considering 
only semi-simple representations of the $\cW_3$ algebra. In this case, eigenvalues 
which differ by integers have to be considered as coinciding in Okubo's classification 
scheme
\footnote{\,
We thank Professor Y. Haraoka for explaining this point to us.
}.
The eigenvalues of the two $\ll 2 \! \times \! 2\rr$ diagonal block matrices $A_{11}$ 
and $A_{22}$ are, 

\begin{equation}
A_{11} \sim \text{diag}  
\ll 1 + 2 b^2, 2 (1 + 2 b^2)\rr, \quad \text{and} \quad  
A_{22}\sim \text{diag} 
\ll 1 + 2 b^2, 2 (1 + 2 b^2)\rr 
\end{equation}

\noindent are associated respectively to the  $\beta_i$ and $\alpha_i$ local 
exponents, $i=1, 2$. In Okubo's classification scheme, our Fuchsian differential 
equation corresponds to

\begin{equation}
n=4, \quad 
q=2, \quad 
m_1 = m_2 = 2, \quad 
p=2, \quad 
n_1 = 2 \quad n_2=1
\end{equation}

\noindent In other words, equation (\ref{DE}) belongs to class-II of accessory-free
Fuchsian systems. This result is interesting, as one would have expected that the 
presence of multiplicities in the adjoint representation implies accessory parameters 
in the Fuchsian equation. For instance in \cite{fl07c} a Fuchsian equation with
accessory parameters was found in the so called \textit{\lq heavy\rq} semi-classical 
limit of the 3-point Toda correlation function. Finally, it is interesting to 
notice that in \cite{bonelli2012vertices} $n$-point conformal block, $n>4$, with
fully-degenerate fields were explicitly given in terms of $_{3}F_{2}$ hypergeometric
functions, that is class-I. It would be interesting to check if, in an appropriate 
limit, one could recover the solutions of the class-II Fuchsian system obtained 
in our case.

\section{Constructing monodromy-invariant function}
\label{section.04.coulomb.gas}

Because we chose the operators such that we have admissible fusion, as implied
in particular in the relation \eqref{alar}, the conformal blocks have the following
Coulomb gas representation, 

\begin{equation}
\left<
\ll  \oint du  V_{b \, \vec{e}_1} (u)  \rr  
\ll  \oint dv  V_{b \, \vec{e}_2} (v)  \rr
V_{   \vec{\alpha}_R                                    } (0) 
V_{-b \, \vec{\omega}_1                                    } (z) 
V_{-b(\vec{\omega}_1 + \vec{\omega}_2)                  } (1) 
V_{ 2Q\vec{\rho} - \vec{\alpha}_R + b \, \vec{\omega}_1} (\infty) 
\right>,
\end{equation}

\noindent One can directly verify that the charges of the vertices entering 
the above formula add up to $2Q\vec{\rho}$, thus satisfying the $\slthree$ 
Coulomb gas neutrality condition \eqref{neutrality} with $n_1=n_2=1$ and 
$m_1=m_2=0$. The above integral yields

\begin{multline}
    z^{b \,  \vec{\omega}_1                   \cdot \vec{\alpha}_R} 
(1-z)^{b \, (\vec{\omega}_1 + \vec{\omega}_2) \cdot \vec{\alpha}_R} 
\\
\oint du \, \oint dv \,   
\ll
    u^{-b \, \vec{e}_1 \cdot \vec{\alpha}_R} 
    v^{-b \, \vec{e}_2 \cdot \vec{\alpha}_R} 
(u-z)^{b^2} 
(u-1)^{b^2}  
(v-1)^{b^2}  
(v-u)^{b^2} 
\rr  
\end{multline}

We need to look for all independent ways of positioning the integration contours. 
Given that our differential equation is fourth-order, we expect four 
linearly-independent such choices. It is convenient to use contours that are open 
paths between singularities. These integrals can be obtained by shrinking the closed 
paths that surround these singularities. In particular, each variable $u$ and $v$ 
is integrated along one of the four paths,

\begin{equation}
\cC_1 : \, -\infty \to       0, \quad
\cC_2 : \,       0 \to       z, \quad
\cC_3 : \,       z \to       1, \quad
\cC_4 : \,       1 \to  \infty
\end{equation}

We need to be careful, as the integration paths of $u$ and $v$ are not allowed to
intersect, and we need to specify which contour lies on top of the other contour. 
There are 16 possible choices, which are illustrated in the following Figure,
 
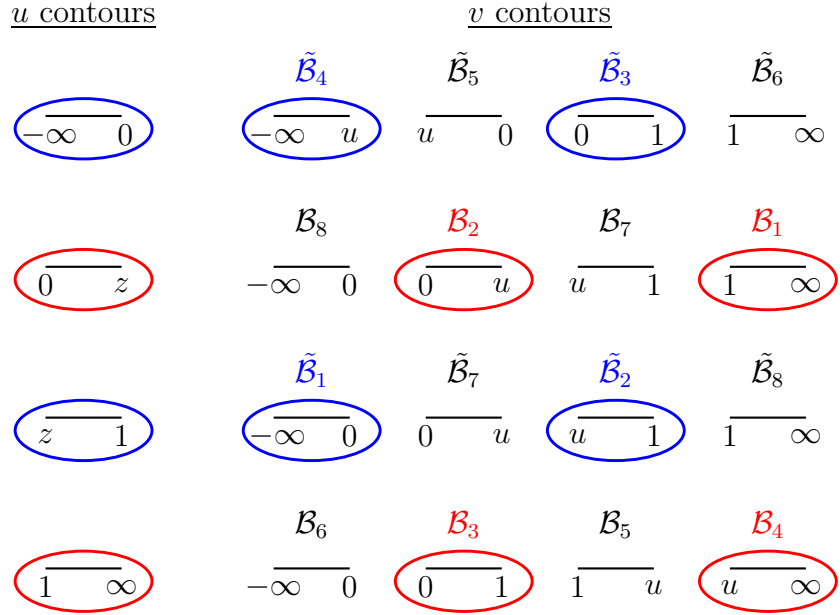
\begin{equation}
\begin{tikzpicture}
[line width=1.1pt]
\draw (0.5,1) node[above]{\underline{$u$ contours}};
\draw (6.5,1) node[above]{\underline{$v$ contours}};
\draw[thick] (0,0)--(1,0);
\draw (0+0.05,0+0.05) node[below]{$-\infty$};
\draw (1+0.05,0+0.05) node[below]{$0$};
\draw[blue] (0.5,0-0.17) ellipse (0.9cm and 0.4cm);
\draw[thick] (3,0)--(4,0);
\draw (3+0.05,0+0.05) node[below]{$-\infty$};
\draw (4,0) node[below]{$u$};
\draw[blue] (3.5,0-0.17) ellipse (0.9cm and 0.4cm);
\draw[blue] (3.5,0.25) node[above]{$\tilde{\cB}_4$};
\draw[thick] (5,0)--(6,0);
\draw (5,0) node[below]{$u$};
\draw (6+0.05,0+0.05) node[below]{$0$};
\draw (5.5,0.25) node[above]{$\tilde{\cB}_5$};
\draw[thick] (7,0)--(8,0);
\draw (7+0.05,0+0.05) node[below]{$0$};
\draw (8+0.05,0+0.05) node[below]{$1$};
\draw[blue] (7.5,0-0.17) ellipse (0.9cm and 0.4cm);
\draw[blue] (7.5,0.25) node[above]{$\tilde{\cB}_3$};
\draw[thick] (9,0)--(10,0);
\draw (9+0.05,0+0.05) node[below]{$1$};
\draw (10+0.05,0) node[below]{$\infty$};
\draw (9.5,0.25) node[above]{$\tilde{\cB}_6$};

\draw[thick] (0,-2)--(1,-2);
\draw (0,-2+0.05) node[below]{$0$};
\draw (1,-2) node[below]{$z$};
\draw[red] (0.5,-2-0.17) ellipse (0.9cm and 0.4cm);
\draw[thick] (3,-2)--(4,-2);
\draw (3,-2+0.05) node[below]{$-\infty$};
\draw (4,-2+0.05) node[below]{$0$};
\draw (3.5,-1.75) node[above]{$\cB_8$};
\draw[thick] (5,-2)--(6,-2);
\draw (5,-2+0.05) node[below]{$0$};
\draw (6,-2) node[below]{$u$};
\draw[red] (5.5,-2-0.17) ellipse (0.9cm and 0.4cm);
\draw[red] (5.5,-1.75) node[above]{$\cB_2$};
\draw[thick] (7,-2)--(8,-2);
\draw (7,-2) node[below]{$u$};
\draw (8,-2+0.05) node[below]{$1$};
\draw (7.5,-1.75) node[above]{$\cB_7$};
\draw[thick] (9,-2)--(10,-2);
\draw (9,-2+0.05) node[below]{$1$};
\draw (10,-2) node[below]{$\infty$};
\draw[red] (9.5,-2-0.17) ellipse (0.9cm and 0.4cm);
\draw[red] (9.5,-1.75) node[above]{$\cB_1$};

\draw[thick] (0,-4)--(1,-4);
\draw (0,-4) node[below]{$z$};
\draw (1,-4+0.05) node[below]{$1$};
\draw[blue] (0.5,-4-0.17) ellipse (0.9cm and 0.4cm);
\draw[thick] (3,-4)--(4,-4);
\draw (3+0.05,-4+0.05) node[below]{$-\infty$};
\draw (4,-4+0.05) node[below]{$0$};
\draw[blue] (3.5,-4-0.17) ellipse (0.9cm and 0.4cm);
\draw[blue] (3.5,-3.75) node[above]{$\tilde{\cB}_1$};
\draw[thick] (5,-4)--(6,-4);
\draw (5,-4+0.05) node[below]{$0$};
\draw (6,-4) node[below]{$u$};
\draw (5.5,-3.75) node[above]{$\tilde{\cB}_7$};
\draw[thick] (7,-4)--(8,-4);
\draw (7,-4) node[below]{$u$};
\draw (8,-4+0.05) node[below]{$1$};
\draw[blue] (7.5,-4-0.17) ellipse (0.9cm and 0.4cm);
\draw[blue] (7.5,-3.75) node[above]{$\tilde{\cB}_2$};
\draw[thick] (9,-4)--(10,-4);
\draw (9,-4+0.05) node[below]{$1$};
\draw (10,-4) node[below]{$\infty$};
\draw (9.5,-3.75) node[above]{$\tilde{\cB}_8$};

\draw[thick] (0,-6)--(1,-6);
\draw (0,-6+0.05) node[below]{$1$};
\draw (1,-6) node[below]{$\infty$};
\draw[red] (0.5,-6-0.17) ellipse (0.9cm and 0.4cm);
\draw[thick] (3,-6)--(4,-6);
\draw (3,-6+0.05) node[below]{$-\infty$};
\draw (4,-6+0.05) node[below]{$0$};
\draw (3.5,-5.75) node[above]{$\cB_6$};
\draw[thick] (5,-6)--(6,-6);
\draw (5,-6+0.05) node[below]{$0$};
\draw (6,-6+0.05) node[below]{$1$};
\draw[red] (5.5,-6-0.17) ellipse (0.9cm and 0.4cm);
\draw[red] (5.5,-5.75) node[above]{$\cB_3$};
\draw[thick] (7,-6)--(8,-6);
\draw (7,-6+0.05) node[below]{$1$};
\draw (8,-6) node[below]{$u$};
\draw (7.5,-5.75) node[above]{$\cB_5$};
\draw[thick] (9,-6)--(10,-6);
\draw (9,-6) node[below]{$u$};
\draw (10,-6) node[below]{$\infty$};
\draw[red] (9.5,-6-0.17) ellipse (0.9cm and 0.4cm);
\draw[red] (9.5,-5.75) node[above]{$\cB_4$};
\end{tikzpicture}
\nonumber
\end{equation}

\vspace{.5cm}

In the above Figure, a function $\mathcal{B}_i$ or $\tilde{\mathcal{B}}_i$,
$i = 1, 2, 3, 4$, is associated to each of the sixteen contour choices. For any of the four domains of integration of the variable $u$, indicated on the left column, there correspond four possible domains of integration of the variable $v$, represented on the same row on the right. For instance 
the $\mathcal{B}_5$ function is given by the integral where $u \in [1,\infty]$ and 
$v \in [1,u]$,

\begin{multline}
\cB_5 (z) =   
          z^{b \,  \vec{\omega}_1                   \cdot \vec{\alpha}_R} 
\ll 1-z \rr^{b \, (\vec{\omega}_1 + \vec{\omega}_2) \cdot \vec{\alpha}_R}  
\\
\int_1^\infty      \! du \,  
\int_1^u \! dv \,  
\ll
u^{-b \, \vec{e}_1 \cdot \vec{\alpha}_R} 
v^{-b \, \vec{e}_2 \cdot \vec{\alpha}_R} 
(z-u)^{b^2} 
(1-u)^{b^2} 
(v-1)^{b^2} 
(v-u)^{b^2}   
\rr\!,
\end{multline} 

\noindent The red (blue) ellipses indicate the $s$- ($t$-) channel
solution basis of the differential equation \eqref{DE}. Let us consider the $s$-channel
that correspond to the  conformal blocks with abelian monodromy around $z=0$. They have
the form $\cB_i(z) = z^{\alpha_i} \sum_{n \geq 0} a^i_n z^n$. Consider for instance the
$\cB_1(z)$ integral,

\begin{multline}
\cB_1 (z) =   
          z^{b \,  \vec{\omega}_1                   \cdot \vec{\alpha}_R} 
\ll 1-z \rr^{b \, (\vec{\omega}_1 + \vec{\omega}_2) \cdot \vec{\alpha}_R}  
\\
\int_0^z      \! du \,  
\int_1^\infty \! dv \,  
\ll
u^{-b \, \vec{e}_1 \cdot \vec{\alpha}_R} 
v^{-b \, \vec{e}_2 \cdot \vec{\alpha}_R} 
(z-u)^{b^2} 
(1-u)^{b^2} 
(v-1)^{b^2} 
(v-u)^{b^2}   
\rr\!,
\end{multline}

\noindent By a simple change of variables, the $\cB_1(z)$ can be written as,  

\begin{multline}
\label{B1}
\cB_1 (z) = 
    z^{\alpha_1} 
\ll 1-z \rr^{b \, (\vec{\omega}_1 +\vec{\omega}_2) \cdot \vec{\alpha}_R}  
\\
\int_0^1      \! du \, 
\int_1^\infty \! dv \, 
\ll
u^{-b \, \vec{e}_1 \cdot \vec{\alpha}_R} 
v^{-b \, \vec{e}_2 \cdot \vec{\alpha}_R} 
  (1-u)^{b^2} 
(1-z u)^{b^2}  
  (v-1)^{b^2}  
 (v-zu)^{b^2}   
\rr\!,
\end{multline}

\noindent that has the correct small $z$ behavior. The fact that $\cB_i$ has a simple
monodromy around $z=0$ can be understood by contour deformation techniques. We have
verified that $\cB_{1}(z)$ and $\cB_{2}(z)$ are solutions of the differential equation
\eqref{DE}. Finally, we have also verified that the functions $\cB_i(z)$, $i=3,4,5,6$, 
are all solutions of the differential equation with local exponents $\alpha_3$, given 
in \eqref{alphas}. For instance, it is easy to verify that,

\begin{multline}
\label{B3}
\cB_3 (z) =   
          z^{b \,   \vec{\omega}_1                  \cdot \vec{\alpha}_R} 
\ll 1-z \rr^{b \, ( \vec{\omega}_1 +\vec{\omega}_2) \cdot \vec{\alpha}_R}  
\\
\int_1^\infty \! du\,  
\int_0^1      \! dv \,  
\ll
u^{-b \, \vec{e}_1 \cdot \vec{\alpha}_R} 
v^{-b \, \vec{e}_2 \cdot \vec{\alpha}_R} 
(u-z)^{b^2} 
(u-1)^{b^2} 
(1-v)^{b^2} 
(u-v)^{b^2}   
\rr\!,
\end{multline}

\noindent behaves for small $z$ as $z^{\alpha_3}(a_0+ a_1 z+\cdots)$. As we discuss 
below, only two functions among the $\cB_i(z)$,  $i=3,4,5,6$ are independent. We choose
arbitrary $\cB_{3}(z)$ and $\cB_{4}(z)$ as the other two functions of the $s$-channel 
basis. The same consideration can be done for the $t$-channel basis, associated to the 
$\beta_i$ local exponents, defined in \eqref{betas}. 

\subsection{The monodromy group}

The monodromy group is generated by the monodromy matrices $\cM_0$ and $\cM_1$.  
In the reference $s$-channel basis

\begin{equation}
\{ \cB_1 (z), \cB_2 (z), \cB_3 (z), \cB_4 (z)\}
\end{equation}
the matrix $\cM_0$ is diagonal:

\begin{equation}
\label{m0}
\cM_0 = \text{diag}
\ll
e^{2 \pi i \alpha_1},
e^{2 \pi i \alpha_2},
e^{2 \pi i \alpha_3},
e^{2 \pi i \alpha_3}
\rr
\end{equation}

\noindent To find the matrix $\cM_1$, we need the $4\times 4$ matrix 
$L_{i,j}$ which relates the $\cB_i$ to the  $\tilde{\cB}_j$,

\begin{equation}
\label{m1t}
\tilde{\cB}_j = L_{j,i} \cB_i
\end{equation}

\noindent Noting that 

\begin{equation}
\label{m1}
\tilde{\cM}_1 = 
\text{diag}
\ll
e^{2 \pi i \alpha_1},
e^{2 \pi i \alpha_2},
e^{2 \pi i \alpha_3},
e^{2 \pi i \alpha_3}
\rr
\end{equation}

\noindent we have,

\begin{equation}
\cM_1 = L^{-1}  \tilde{\cM}_1 L
\end{equation}

We could find the matrix $L$ explicitly by applying the residue theorem systematically. 
For instance, fix the variable $u \in (0,z)$ as a parameter and consider the $v$ 
integrals. We denote

\begin{equation}
\label{xy}
	x = - b \, \vec{e}_1 \cdot \vec{\alpha}_R\;, \quad y = -b \, \vec{e}_2 \cdot \vec{\alpha}_R\;.
\end{equation}

\noindent Fixing $u$ and moving the variable $v$, Cauchy theorem provides relations of the kind

\begin{multline}
  \ll \int_{-\infty}^0 \! dv ... \rr 
+ e^{\pm i \pi  y         } \ll \int_0^u         \! dv ... \rr 
+ e^{\pm i \pi (y +   b^2)} \ll \int_u^1         \! dv ... \rr 
\\
+ e^{\pm i \pi (y + 2 b^2)} \ll \int_1^\infty  \! dv ... \rr=0 .
\end{multline}

\noindent Using these, one can find a linear relation between the contour integrals, 
such as 

\begin{equation}
\sin{\ll \pi      b^2 \rr} \, \cB_2(z) + 
\sin{\ll \pi (y+  b^2)\rr} \, \cB_7(z) + 
\sin{\ll \pi (y+2 b^2)\rr} \, \cB_1(z) = 0.
\end{equation}

\noindent Similarly, by fixing $v$ and moving the variable $u$, we obtain relation between 
$\cB_i$ and $\tilde{\cB}_j$ functions, such as

\begin{multline}
\sin{\ll \pi       b^2 \rr} \, \tilde{\cB}_4(z) + 
\sin{\ll \pi (x+   b^2)\rr} \,       {\cB}_8(z) + 
\sin{\ll \pi (x+ 2 b^2)\rr} \, \tilde{\cB}_1(z) 
\\
+ \sin{\ll \pi (x+ 3 b^2)\rr} \,       {\cB}_6(z) = 0
\end{multline} 

\noindent We find out $12$ independent relations which allow 
to write the matrix $L$  (using short notation $[x]=\sin(\pi x)$)  as follows
 
\begin{align}
\label{monodromyL}
L=\resizebox*{0.6\vsize}{!}{$ 
\ll
\begin{array}{cccc}
\textsc{-} \frac{[b^2]     [x    ]               }{[b^2+x] [b^2+y]              } &
           \frac{[b^2]     [b^2+x+y]               }{[b^2+y] [2 b^2+x+y]          } &
           \frac{[b^2]     [3 b^2+x] [  b^2+x+y]     }{[b^2+x] [b^2+y] [2 b^2+x+y]}     & 
\textsc{-} \frac{[b^2]     [x    ] [4 b^2+x+y]     }{[b^2+x] [b^2+y] [2 b^2+x+y]} 
\\
\frac{[x]     [2 b^2+y]                          }{[b^2+x] [b^2+y]              } &
\frac{[y]     [b^2+x+y]                          }{[b^2+y] [2 b^2+x+y]          } & 
\textsc{-}  \frac{[3 b^2+x] [y    ] [3 b^2+x+y]    }{[b^2+x] [b^2+y] [2 b^2+x+y]    } &  
\textsc{-}  \frac{[2b^2+x]  [2 b^2+y] [4 b^2+x+y]    }{[b^2+x] [b^2+y] [2 b^2+x+y]    } 
\\
            \frac{[b^2]     [2 b^2+y]              }{[b^2+x] [b^2+y]              } &  
\textsc{-}  \frac{[b^2]^2                        }{[b^2+y] [2 b^2+x+y]          } & 
            \frac{[b^2+y]   [b^2+x+y]+ [y] [4 b^2+x+y]}{[b^2]^{-1} [b^2+x] [b^2+y] [2 b^2+x+y] } &
            \frac{[b^2] [2 b^2+y] [4 b^2+x+y]         }{[b^2+x] [b^2+y] [2 b^2+x+y]          } 
\\
\textsc{-}  \frac{[b^2]                                 }{[b^2+x] [b^2+y]} & 
\textsc{- } \frac{[b^2] [y]                                 }{[b^2+y] [2 b^2+x+y]} &  
\textsc{- } \frac{[b^2] [3 b^2+x] [y]                    }{[b^2+x] [b^2+y] [2 b^2+x+y]} & 
\textsc{-}  \frac{[x][b^2+y] + [3 b^2+x] [2 b^2+y]         }{[b^2]^{-1} [b^2+x] [b^2+y] [2 b^2+x+y]} 
\end{array}
\rr 
$}.
\end{align}

\subsection{Monodromy-invariant correlation function}

We consider the general 4-point correlation function

\begin{equation}
	G(z, \bar{z}) = \cB(z)_T X \cB (\bar{z}),
\end{equation} 
where $\cB(z) =(\cB_{1}(z),\cB_{2}(z),\cB_{3}(z),\cB_{4}(z))$ is the array of 
$s$-channel solutions, $\cB(z)_T$ its transpose and $X$ is a $4\times 4$ matrix 
of coefficients which do not depend on $z$. We look for constraints on $X_{i, j}$, 
imposed by demanding

\begin{equation}
\label{monodromies}
\cM_0 G(z, \bar{z}) = G(z, \bar{z}) \quad 
\cM_1 G(z, \bar{z}) = G(z, \bar{z}) 
\end{equation}

\noindent From the first relation in \eqref{monodromies} and from \eqref{m0}, the matrix $X$ has to have the following form

\begin{align}
X=
\ll
\begin{array}{cccc}
x_1 & 0 & 0 & 0  \\
 0 & x_2 & 0 & 0 \\
 0 & 0 & x_3 & x_5 + i\; x_6 \\
0 & 0 & x_5- i\; x_6 & x_4
\end{array}
\rr\;,
\end{align}

\noindent Up to a  global normalization, we have five unknown constants $x_i/x_1$, 
$i=2,\cdots,6$ to determine. This will be fixed by imposing the second constraint 
in \eqref{monodromies}. In order to know the action of $\cM_1$ on $G$, we write 
the function $G$ in terms of the $t$-channel basis 
$\tilde{\cB} (z) = \ll \tilde{\cB}_1(z), \tilde{\cB}_2(z), \tilde{\cB}_3(z), \tilde{\cB}_4 (z) \rr$

\begin{equation}
G(z, \bar{z}) = \tilde{\cB}(z)_T Y \tilde{\cB} (\bar{z}),
\end{equation}

\noindent In term of the monodromy matrix $L$, given in \eqref{monodromyL}, the matrix 
$Y$ is

\begin{equation}
Y =  \ll L^{-1} \rr^T X L^{-1}
\end{equation}

\noindent Taking \eqref{m1} into account, the relation $\cM_1 G(z, \bar{z}) = G(z, \bar{z})$ 
implies five equations:

\begin{equation}
\label{systmon}
Y_{1, 2} = 0, \quad Y_{1, 3} = 0,\quad  Y_{1, 4} = 0,\quad Y_{2, 3} = 0,\quad Y_{2, 4} = 0.
\end{equation}

\noindent The above equation form a system of linear equations in the coefficients $x_i$.  
We have checked that the five relations are independent. The system \eqref{systmon} has 
rank five and allows to fix the five rations $x_i/x_1$, $i=2,\cdots,6$. We note that  
$x_i^{\text{sol}}$, $i=1,\cdots,6$ represent the solutions of the above system. 
The value $x^{\text{sol}}_{1}$ can be chosen arbitrary to fix the global normalization 
of the correlation $G(z, \bar{z})$. We found $x_{6}^{\text{sol}}=0$. The other solutions 
$x_i^{\text{sol}}$, $i=2,\cdots,5$, which depend on $x,y$ and $b$ (or equivalently on 
$a_{R_1}$,$a_{R_1}$ and $b$ see \eqref{xy}), are given in terms of trigonometric functions. 
It is convenient to express $G(z, \bar{z})$ in terms of the normalized functions,

\begin{equation}
\cB^{\text{norm}}_i(z) = \frac{\cB_i(z)}{\cB_i(0)},
\end{equation}
where:
\begin{eqnarray}
\cB_1(0) & = & \frac{\Gamma(1 + b^2)^2 \Gamma(1 + x) \Gamma(-1 - 2 b^2 - y)
}{
\Gamma(2 + b^2 + x) \Gamma(-b^2 - y)}, 
\\
\cB_2(0) & = & \frac{\Gamma(1 + b^2)^2 \Gamma(1 + y) \Gamma(2+ b^2 +x+y)
}{
\Gamma(2 + b^2 + y) \Gamma(3+2 b^2 +x+y)}, 
\\
\cB_3(0) & = & \frac{\Gamma(1 + b^2)^2 \Gamma(-1 - 3 b^2 - x)
\Gamma(1 + y)}{\Gamma(-2 b^2 - x) \Gamma(2 + b^2 + y)}\; {} \times 
\\
& & \times \; {}_3F_{2}(-b^2, -1 - 3 b^2 - x, 1 + y;-2 b^2 - x, 2 + b^2 + y|1), 
\nonumber 
\end{eqnarray}

\begin{eqnarray}
\cB_4(0) & = & \frac{\Gamma(1 + b^2)^2 \Gamma(-1 - 2 b^2 - y)
\Gamma(-2 -4b^2-x-y)}{\Gamma(-b^2 - y) \Gamma(-1 -3b^2 -x-y)}\; {} \times  
\\
& & \times \; _3F_{2}(-b^2, -1 - 2 b^2 - y, -2 -4b^2 -x-y;-b^2 - y, -1 -3 b^2 -x-y|1). 
\nonumber 
\label{norms}
\end{eqnarray}

\noindent Setting $x_{1}^{\text{sol}}=\cB_{1}(0)^2$ and

\begin{equation}
\label{Xx}
X_i = \frac{\cB_i(0)^2}{\cB_{1}(0)^2} \;x_i^{sol},
\end{equation}

\noindent the monodromy invariant correlation function takes the form,

\begin{multline}
 G(z, \bar{z}) =  \cB^{\text{norm}}_1(z)\cB^{\text{norm}}_1(\bar{z})+ 
 X_{2}\cB^{\text{norm}}_2(z)\cB^{\text{norm}}_2(\bar{z})+ X_{3}
 \cB^{\text{norm}}_3(z)\cB^{\text{norm}}_3(\bar{z}) 
 \\ 
 +  X_4 \cB^{\text{norm}}_4 (z) \cB^{\text{norm}}_4 (\bar{z}) + 
 X_5 \ll\cB^{\text{norm}}_3 (z)\cB^{\text{norm}}_4 (\bar{z}) + 
 \cB^{\text{norm}}_4 (z) \cB^{\text{norm}}_3 (\bar{z})
 \rr\!\!.
\end{multline}

\noindent The function $G(z,\bar{z})$ have therefore the following small 
$z$ behavior,

\begin{multline}
 G(z, \bar{z}) = |z|^{2\alpha_1} \ll 1 + \cO(z, \bar{z}) \rr + 
 X_2 |z|^{2 \alpha_2}\ll 1
 + \cO (z, \bar{z}) \rr
 \\
 + (X_3 + 2X_5 + X_4)|z|^{-2\alpha_3}\ll 1 + \cO(z, \bar{z}) \rr\!\!,
 \end{multline}
 
\noindent where the exponents have been defined in \eqref{alphas} and correspond to 
the channels \eqref{ch1}, \eqref{ch2} and \eqref{ch34}. We can now directly compare 
our findings with the structure constants computed in \cite{fl07c} and re-derived here 
in \ref{Structconst}. In the second channel, we found that,

\begin{equation}
X_{2} = 
\frac{
\gamma(-x) \gamma(2+x+b^2) \gamma(-y-b^2) \gamma(2+y+2b^2)
}{
\gamma(-y) \gamma(2+y+b^2) \gamma(-1-x-y-b^2) \gamma(3+x+y+2b^2)}.
\end{equation}

\noindent Noting as $\vec{\alpha}_{M}^{(1)}$ and $\vec{\alpha}_M^{(2)}$ the charges
$\vec{\alpha}_M$ corresponding respectively to \eqref{ch1} and to \eqref{ch2}, one 
can directly verify that,

\begin{equation}
X_2 = 
\frac{
C_{-b \vec{\omega}_1, \vec{\alpha}_R}^{\vec{\alpha}^{(2)}_M} 
C_{-b \vec{\rho},     \vec{\alpha}^{(2)}_M}^{\vec{\alpha}_L}
}{
C_{-b\vec{\omega}_1,  \vec{\alpha}_R}^{\vec{\alpha}^{(1)}_M} 
C_{-b\vec{\rho},      \vec{\alpha}_M^{(1)}}^{\vec{\alpha}_L}
},
\label{X2}
\end{equation}

\noindent where we recall that the $\vec{\alpha}_L$ is related to $\vec{\alpha}_R$ 
by the \eqref{alar}. The structure constants $C_{-b\vec{\omega}_1, 
\vec{\alpha}_R}^{\vec{\alpha}^{(i)}_M}$ and $C_{-b\vec{\rho}, 
\vec{\alpha}^{(i)}_M}^{\vec{\alpha}_L}$ are given in \eqref{1node} and in \eqref{2node}. Note that the normalization of the fields $\Phi_{\vec{\alpha}_L}$ and $\Phi_{\vec{\alpha}_R}$ are irrelevant in this ratio, and therefore we can use the vertex operators to compute these three point functions. 

More interesting is the channel \eqref{ch34}, $\vec{\alpha}_M = \vec{\alpha}_L$ 
associated with the constants $X_3$, $X_4$ and $X_5$. We recall that the existence 
of a two-dimensional space of solutions spanned by $\cB_{3}(z)$ and $\cB_{4}(z)$ is 
related to the two-fold multiplicity of the $0$ weight in the adjoint representation.  
In the OPE between $\Phi_{2211}$ and $\Phi_L$ this ambiguity can be expressed by 
a free parameter $\iota$ appearing from the first-order of the expansion,

\begin{multline}
\Phi_{2211}(z) 
\Phi_{L}(0) = 
\\
z^{-h} \Phi_L (0) + z^{-h+1} \ll \frac{h}{2h_L} L_{-1} + \iota 
\left(
-3 q_L L_{-1} + 2 h_L W_{-1} \right) \rr \Phi_L (0)+ \cO (z^{-h+2}).
\end{multline}

\noindent Any conformal block obtained by the linear combination $\cB_{3}(z)+ s(\iota)
\cB_{4}(z)$, which is therefore also a solution of the differential equation \eqref{DE}, 
is associated to a particular value of $\iota$. Note that in the above expansion we 
have chosen a particular basis of state at the first level, namely $L_{-1} \Phi_L$ 
and $\chi(z)\equiv \left(-3 q_L L_{-1} + 2 h_L W_{-1} \right)\Phi_L (z)$.
This choice can be considered as the most natural as $\chi(z)$ is a \textit{Virasoro} 
primary field (it is $\cW_3$ primary only if $\Phi_L$ is associated to the 
$\vec{\omega}_1$ and $\vec{\omega}_2$ representation). 
We have verified that

\begin{equation}
X_3 + 2 X_5 + X_4 = 
\frac{ 
C_{-b \vec{\omega}_1, \vec{\alpha}_R}^{\vec{\alpha}_L} 
C_{-b\vec{\rho}, \vec{\alpha}_L}^{\vec{\alpha}_L}
}{
C_{-b\vec{\omega}_1, 
\vec{\alpha}_R}^{\vec{\alpha}^{(1)}_M} C_{-b\vec{\rho}, 
\vec{\alpha}_M^{(1)}}^{\vec{\alpha}_L}},
\end{equation} 

\noindent where the structure constant 
$C_{-b\vec{\rho}, \vec{\alpha}_L}^{\vec{\alpha}_L}$ has been derived in \eqref{intdiff}.

The above result shows that the monodromy requirement eliminates the ambiguities due 
to the presence of multiplicities. This is of course expected since $G(z, \bar{z})$ 
has a Coulomb gas representation of type \eqref{CGcomp}, which has a simple monodromy 
by construction. On the other hand, the Coulomb gas approach of type \eqref{CGhol} 
shows in the most transparent way how the monodromy solves the problem of ambiguities. 
Moreover, we have computed all the constants $X_i$ in \eqref{Xx} from the solutions 
of the linear system \eqref{systmon} and from the values of the norms \eqref{norms}. 
We stress that these constants are hardly accessible \textit{via} the study of the 
Coulomb gas representation \eqref{CGcomp} of $G(z,\bar{z})$, while the direct 
computation of the 3-point function \textit{via} the Coulomb gas gives access only 
to a particular combination of these three values.

\section{Summary and discussion}
\label{section.06.summary.comments}

In this paper we studied $\cW_3$ Toda conformal field theory. We considered the explicit 
construction of matrix elements, conformal blocks and correlation functions of primary 
fields in the case where the primary fields do not belong to the special class that is 
available to the AGT approach \cite{agt09, wyl09}. We recall here that, in general, 
$\cW_N$ AGT allows one to construct matrix elements for the fields with highest-weights
proportional either $\omega_1$ or $\omega_{N-1}$ fundamental weights of $\slN$. In this 
case, the correspondence between 2-dimensional conformal field theory and 4-dimensional 
supersymmetric gauge theories, as proposed in \cite{agt09}, is available, and $\cW_N$ 
conformal blocks are equal to Nekrasov instanton partition functions \cite{wyl09}. The 
core of the problem of extending these results is that for $N \geq 3$ the general matrix 
element of a primary field between two descendant states is not expressed solely in terms 
of the primary 3-point function 
but involves also an infinite set of new independent basic matrix elements, so to say,  
$\cW_N$-partners. For the special class of semi-degenerate fields this problem can be 
resolved and all matrix elements, as well as the coefficients of the operator product 
expansion, can be fixed uniquely in terms of 3-point functions of primary fields 
\footnote{\,
See \cite{bonelli2012vertices, gl14, pom14I, pom14II, furlan2015some} for recent works
towards a more general analysis.
}. 

In this paper, we focused on the simplest example of non-AGT type: a primary fully-degenerate
field in the adjoint representation of $\slthree$ algebra in $\cW_3$ Toda theory. What we call non-AGT-type refers strictly to the case in which the linear identities of type (\ref{Level.1.identity}) are not sufficient, and one needs the non-linear identity of type (\ref{level.2.identity}).   We 
showed how null-state conditions in this case allow one to determine the operator 
product expansion, and, in particular, its first level coefficients, which, as one might
think naively, could remain unfixed. This case is of particular interest since it contains
non-trivial effects of the multiplicity problem. Using representation theory, we showed
that these matrix elements can be computed explicitly. 

Next, we constructed $4$-point conformal blocks which involves matrix elements of the 
new type. We checked our results against the differential equation which follows from 
the null-state conditions for one semi-degenerate fundamental and one fully-degenerate 
adjoint fields.  
We found the Okubo form of the obtained fourth-order Fuchsian differential equation, 
and showed that it is an accessory-free Fuchsian system of the class-II. We solved 
the differential equation in terms of Coulomb gas integrals and computed the monodromy 
group. Finally, we showed that in the presence of multiplicity at the level of the matrix
elements, using the bootstrap technique, the correlation functions can be constructed 
uniquely and do not contain arbitrary parameters. As a byproduct, we obtained explicit 
expressions for the particular structure constants that are not accessible by means of 
the Coulomb gas representation. 
 
A natural question for further study is the generalisation of the discussed methods 
for more general classes of non-fundamental or anti-fundamental fields. In this context, 
constructing semi-degenerate fields in higher representations of $\slN$ seems to be
relevant. Work on this problem is in progress \cite{semideg}. Another interesting
question is the possible extension  of rational and non-rational versions of $\cW_N$  
AGT \cite{wyl09, bfs15, Alkalaev:2014sma, Bershtein:2014sma} on the class of non-fundamental 
fields.
This problem requires constructing vertex operators in the composite $\cW_N \times \cH$
conformal field theory, where $\cH$ is the Heisenberg algebra. In this context, a search
for the \textit{dressing} Heisenberg field for the non-fundamental fields $\cW_N$ is
required.
 
In \cite{pom14I, pom14II}, general $\cW_N$ 3-point functions are computed, starting 
from topological string partition functions. Strictly speaking, these results are in 
the context of a $q$-deformed version of a $\cW_N \times \cH$ conformal field theory, 
where $\cH$ is the Heisenberg algebra. The $q \rightarrow 1$ limit of these results is, 
at this stage, not entirely straightforward. In \cite{gl14}, an approach towards
$q$-deformed $\cW_N \times \cH$ $(n+2)$-point conformal blocks that involve $\cW_N$ 
primary fields, in representations of $\slN$ that are higher than the fundamental, 
was briefly proposed. The basic idea in this proposal is to take suitable limits of
conformal blocks of primary fields in the fundamental of $\slN$. It remains to carry 
out this proposal in detail. 

\appendix

\section{Shapovalov matrices, at level-one and level-two}
\label{appendix.A.Shapovalov}
We give in the following the Shapovalov matrices, at level-one and level-two, associated to the primary field $\Phi_{M}$ characterized by the $h_M,q_M$ quantum numbers. We use here the notation introduced in \ref{W3Verma}. 
\subsection*{The level-one block}

There are two states at level-one, $I_1=\{1;\emptyset\}$ and $I_2=\{\emptyset; 1\}$. 
In the basis $\ll I_1,I_2 \rr$, the matrix $H_1$ is 

\begin{equation}
H_1=
\ll
\begin{array}{cc}
 2 h_M & 3 q_M \\
 3 q_M & \frac23 
\ll
h_M^2+\frac{h_M}{5}
\rr
- \frac{1}{240} 
(5 c+22) h_M 
\\
\end{array}
\rr
\end{equation}

\subsection*{The level-two block}

There are five states at level-two. 
$I_1=\{2;\emptyset\}$, 
$I_2=\{1,1;\emptyset\}$, 
$I_3=\{\emptyset;2\}$, 
$I_4=\{\emptyset; 1,1\}$, and 
$I_5=\{1;1\}$. 
In the basis $ \ll I_1, I_2, I_3, I_4, I_5 \rr$, we have the $5 \! \times \! 5$ matrix

\begin{multline}
H_2 =
\resizebox{.75 \hsize}{!}
{$
\ll\!
\begin{array}{ccccc}
\frac{c}{2} + 4 \Delta _M                      & 
\phantom{00} &
6 h_M                                     &  
\phantom{00}&
6 q_M                                          
\\
6 \Delta _M                                      & 
\phantom{000}&
4 h_M  \ll 2 h_M+1 \rr                       & 
\phantom{00}&
12 q_M                                            
\\
 6 q_M                                            & 
 \phantom{00}&
12 q_M                                            & 
\phantom{00}&
\frac43 h_M^2 + \ll c+6 \rr  \frac{h_M}{6} 
\\
 \ll 32 h_M+2-c \rr  \frac{5h_M }{48}                                                &
\phantom{00}&
18 q_M^2+4 h_M^2 +  \ll 2-c \rr \frac{h_M}{8} & 
\phantom{00}&
\ll c+14 \rr \frac{q_M}{8} +6 h_M q_M                                                          
\\
 9 q_M                                               & 
\phantom{000} &
12 h_M q_M+6 q_M                               & 
\phantom{00}&
 \ll -c+32 h_M+2 \rr  \frac{h_M}{12}      
 \\
\end{array}
\right.$
}
\\
\resizebox{.98\hsize}{!}
{$
\left.
\begin{array}{ccc}
 - \frac{5}{48}  \ll c-32 h_M - 2 \rr h_M   &  
 \phantom{0000}&
9 q_M 
\\
18 q_M^2 + 4 h_M^2 + \ll 2-c \rr \frac{h_M}{8} & 
\phantom{000}&
12 h_M q_M + 6 q_M 
\\
\ll c+14 \rr \frac{q_M}{8}  + 6 h_M q_M                & 
\phantom{0000}&
 \ll 32 h_M+2 -c\rr \!\!\frac{h_M}{12}
\\
 \ll c^2 - 52 c + 612 \rr  \!\!\frac{h_M^2}{1152} + \!
 \ll c^2 - 36 c +   68 \rr\!\! \frac{h_M    }{2304} +
 \frac{16h_M^4+(18-c)\Delta _M^3+216 q_M^2 }{18}
 & 
\phantom{0000}&
  \ll 32 h_M + 2 -c\rr\!\!   \frac{(2 h_M + 3)q_M}{16}
\\
 \ll 32 h_M + 2 -c \rr\!\!  \frac{(2 h_M + 3)q_M}{16}& 
\phantom{0000}&
9 q_M^2 + \! \ll 32 \Delta _M+2-c \rr  \!\! \frac{(h_M+1)h_M}{24}
\end{array}
\!\rr
$}
\end{multline}

\section{Matrix elements, at level-one and at level-two}
\label{appendix.B.matrix.elements}
We give here the expansion \eqref{cb.series.expansion} till  the second-order of 
a general  $\cW_3$ conformal block. In the following we define

\begin{equation}
W_{1}^{a}\equiv \frac{\langle 
\Phi_{M} 
\, | \, 
\ll W_{-1} \Phi_{1}
\rr 
\, | \,
\Phi_R
\rangle}{\langle 
\Phi_{M} 
\, | \, 
 \Phi_{1}
\, | \,
\Phi_R
\rangle} , \quad 
W_{1}^{b}\equiv \frac{\langle 
\Phi_{L}^{*} 
\, | \, 
\ll W_{-1} \Phi_{2}
\rr 
\, | \,
\Phi_M
\rangle}{\langle 
\Phi_{L}^* 
\, | \, 
 \Phi_{2}
\, | \,
\Phi_M
\rangle} 
\end{equation}
and 
\begin{equation}
W_{1,1}^{a}\equiv \frac{\langle 
\Phi_{M} 
\, | \, 
\ll W_{-1}^2 \Phi_{1}
\rr 
\, | \,
\Phi_R
\rangle}{\langle 
\Phi_{M} 
\, | \, 
 \Phi_{1}
\, | \,
\Phi_R
\rangle} , \quad 
W_{1,1}^{b}\equiv \frac{\langle 
\Phi_{L}^{*} 
\, | \, 
\ll W_{-1}^2 \Phi_{2}
\rr 
\, | \,
\Phi_M
\rangle}{\langle 
\Phi_{L}^* 
\, | \, 
 \Phi_{2}
\, | \,
\Phi_M
\rangle} 
\end{equation}
These elements are unknown for general field $\Phi_{1}$ and $\Phi_{2}$. 
\subsection*{The level-one matrix elements} 

At first-order in the expansion of \eqref{cb.series.expansion}, we have 
the matrix elements

\begin{align}
\Gamma'_{\{1        ; \emptyset\}, \emptyset, \emptyset} & = h_1+h_M-h_R&
	\\
\Gamma'_{\{\emptyset;  1       \}, \emptyset, \emptyset} & = q_M-q_R+2 w_1 + W_1^a&
	\\
\Gamma_{   \emptyset, \emptyset, \{1;\emptyset\}}        & = h_M+h_2- h_L&
	\\
\Gamma_{   \emptyset, \emptyset, \{\emptyset;1\}}        & = q_M+q_L+  q_2 + W^b_1&
\end{align}

\subsection*{ The Level-two matrix elements}
At the second-order in the expansion \eqref{cb.series.expansion}, we have,

\begin{equation}
\Gamma'_{\{2;\emptyset\},\emptyset,\emptyset}  = 2h_1+h_M-h_R
\end{equation}

\begin{equation}
\Gamma'_{\{1,1;\emptyset\},\emptyset,\emptyset}  = (h _1+1+h_M-h_R)(h _1+h_M-h_R)
\end{equation}

\begin{equation}
\Gamma'_{\{\emptyset;2\},\emptyset,\emptyset}    = q_M-w_R+5 q_1 + 2 W_1^a
\end{equation}

\begin{multline}
\Gamma'_{\{\emptyset;1,1\},\emptyset,\emptyset}  
= (q_M-q_R+2 q_1)(q_M-q_R+2 w_1 + 2 W_1^a) +
\\
+2(2/15-\eta/5+2/3 h_1)(h_M-h_1-h_R)+
\\
+ (2/15-\eta/5+2/3 h_1)(h_M+h_1-h_R)+
\\
+ (2/15-\eta/5        ) h_1 + 2/3h_1^2 + W^a_{1,1}
\end{multline}

\begin{equation}
\Gamma'_{\{1;1\},\emptyset,\emptyset}  = (h_M+h_1+1-h_R)(q_M-q_R+2 w_1 +  W_1^a)
\end{equation}

\begin{equation}
\Gamma_{\emptyset,\emptyset,\{2;\emptyset\}}  =    h_M+2h_2- h_L
\end{equation}

\begin{equation}
\Gamma_{\emptyset,\emptyset,\{1,1;\emptyset\}}  = (h_M+h_2+1- h_L)(h_M+h_2- h_L)
\end{equation}

\begin{equation}
\Gamma_{\emptyset,\emptyset,\{\emptyset;2\}}  = q_M+q_L+q_2+ 2 W^b_1
\end{equation}

\begin{multline}
\Gamma_{\emptyset,\emptyset,\{\emptyset; 1,1\}}  = (q_M+q_L+q_2)(q_M+q_L+q_2 + 2 W^b_1) 
\\
+(2/15-\eta/5+2/3 h_2)(h_L-h_M-h_2)+
\\
+(2/15-\eta/5+2/3 h_M)(h_M+h_2-h_L) + W^b_{1,1}
\end{multline}

\noindent In the above expressions, the constant $\eta$ has been defined in \eqref{norm_w}. 
Using the above formula and the expansion \eqref{cb.series.expansion}, the first two orders
in the expansion of a general conformal block can be given in terms of the central charge 
$c$, of the external fields parameters $\{h_L, q_L, h_2, q_2, h_1, q_1, h_R, q_R\}$, of
the fusion channel ones $\{h_M, q_M\}$ and of the four elements $W_1^{a, b}$ and 
$W_{1, 1}^{a,b}$.

\section{$\cW_3$ highest-weight modules with null-states at level-two}
\label{appendix.C.null.states}

We consider a primary field 
$\Phi_{2211}$  and we want to derive the form of the two $\cW_3$ null-states at level-two.  
In general case there are five states $ \ll I_1, I_2, I_3, I_4, I_5 \rr$ at level two 
(see \ref{appendix.A.Shapovalov}). Level-two state 

\begin{equation}
\label{woeigen} 
\Psi(\lambda) = 
\ll 
c_1 (\lambda) L_{-2} +
c_2 (\lambda) W_{-2}   +
c_3 (\lambda) L_{-1}^2 + 
c_4 (\lambda) L_{-1} W_{-1} + 
                     W_{-1}^2 
\rr 
\Phi_{h, q}  
\end{equation} 
is an eigenstate of the operator $W_{0}$ of eigenvalue $\lambda+ q$, if
\begin{align}
c_1 (\lambda) &= \frac{1}{128 \eta^2}       \ll \lambda^4 -(8 f  +16 \eta^2)\lambda^2+ 64 \eta^4  \rr,
\\
c_2 (\lambda) &= \frac{\lambda}{ 32 \eta^2} \ll \lambda^2 - 8 (\eta^2+f)                          \rr, 
\\
c_3 (\lambda) &= \frac18                    \ll \lambda^2 - 8  \eta^2-4 f                         \rr,
\\
c_4 (\lambda) &= \frac{\lambda}{2} ,
\end{align}
where
\begin{equation}
f= -\frac{1}{5} +2 \eta^2 \ll h+\frac15 \rr,
\end{equation}
and $\eta$ is given in \eqref{norm_w}.
 For $h=h_{2211}$ one can verify that the two states \eqref{woeigen}, 
 with $\lambda$ given by
\begin{equation}
\lambda_{\pm} = \pm 2(1+2 h)\sqrt{\frac{2}{(3-h)(1+5h)}}, 
\end{equation}

\noindent are $\cW_3$ primaries. Indeed, the states $\Psi^{\pm}\equiv \Psi(\lambda_{\pm})$, 
are eigenvalues of $W_0$ and obey $L_{+1}\Psi^{\pm} = L_{+2}\Psi^{\pm}=0$. We therefore 
have the two null-state conditions,

\begin{equation}
\Psi_{+} = 0 \quad \Psi_{-} = 0,
\end{equation}
which give the relations \eqref{deg2211} and \eqref{level.2.identity}.

\section{Structure constants from the $\slthree$ Coulomb gas}
\label{Structconst}

We review the computation of the structure constants \eqref{threepoint} that enter in the correlation function \eqref{combdiag} and that can be computed  through Coulomb gas, as done in \cite{fl07c}. 

\subsection{The first node}

The first node of the diagram \eqref{combdiag} corresponds to
\begin{equation}
\begin{tikzpicture}
[line width=1.2pt]
\draw (.3,.5)node[right]{$\vec{\alpha}_M = \vec{\alpha}_R - b \vec{h}_i$};
\draw (-1,1.2) node [left] {$\vec{\alpha}_{2111}$} -- (0,0) -- node[]{}(1.6,0);
\draw (-1,-1.2) node [left] {$\vec{\alpha}_{R}$} -- (0,0);
\end{tikzpicture}
\end{equation}
where $\vec{h}_i$, $i=1,2,3$ are defined in \eqref{hi}.
The structure constant $C_{-b \vec{\omega}_1, \vec{\alpha}_R}^{\vec{\alpha}_R-b \vec{h}_i}$ is defined through the Coulomb gas three point function:
\begin{equation}
C_{-b \vec{\omega}_1, \vec{\alpha}_R}^{\vec{\alpha}_M} =\left<V_{2\vec{\alpha}_0 - \vec{\alpha}_M}(\infty) V_{-b \vec{\omega}_1}(1)V_{\vec{\alpha}_R}(0)\right>.
\end{equation} 
From the neutrality condition \eqref{neutrality}, the Coulomb gas representation of $C_{-b \vec{\omega}_1, \vec{\alpha}_R}^{\vec{\alpha}_R-b \vec{h}_i}$ gives the following representation:
\begin{eqnarray}
C_{-b \vec{\omega}_1, \vec{\alpha}_R}^{\vec{\alpha}_R-b \vec{h}_1} &=& 1 \quad (n_1=n_2=m_1=m_2=0)\nonumber \\
C_{-b \vec{\omega}_1, \vec{\alpha}_R}^{\vec{\alpha}_R-b \vec{h}_2}&=& \int d^2 t |t|^{2 b^2}|t-1|^{2 x} \quad (n_1=1, n_2=m_1=m_2=0) \nonumber \\
C_{-b \vec{\omega}_1, \vec{\alpha}_R}^{\vec{\alpha}_R-b \vec{h}_3}&=& \int d^2 t_1 \int d^2 t_2 |t_1|^{2 b^2}|t_1-1|^{2 x } |t_2|^{0}|t_2-1|^{2 y } |t_1-t_2|^{2 b^2}\nonumber \\
 && \quad \quad \quad \quad (n_1=1, n_2=1, m_1=m_2=0)\nonumber \\
&&
\end{eqnarray}
where $x$ and $y$ are defined in \eqref{xy}. Using the formula (see for instance Appendix 2 of chapter 7 in \cite{dot98}):
\begin{equation}
\label{dotint}
\int d^2 t |t|^{2 a}|t-1|^{2 b} =\pi \frac{\gamma(1+a)\gamma(1+b)}{2+a+b}
\end{equation}  
one derives:
\begin{eqnarray}
C_{-b \vec{\omega}_1} &=& 1 \nonumber \\
C_{-b \vec{\omega}_1, \vec{\alpha}_R}^{\vec{\alpha}_R-b \vec{h}_2}&=& \pi \; \frac{\gamma(1+b^2)\gamma(1+x)}{\gamma(2+b^2+x)}\nonumber  \\
C_{-b \vec{\omega}_1, \vec{\alpha}_R}^{\vec{\alpha}_R-b \vec{h}_3}&=& \pi^2\; \frac{\gamma(1+b^2)^2\gamma(1+y)\gamma(2+b^2+x+y)}{\gamma(2+b^2+y)\gamma(3+2b^2+y+x)}
\label{1node}
\end{eqnarray}

\noindent The above results coincide with those of Equation (1.51) in \cite{fl07c}.

\subsection{The second node}
The second node of the \eqref{combdiag} is illustrated here:
\begin{equation}
\begin{tikzpicture}
[line width=1.2pt]
\draw (1.8,.5)node[right]{$\vec{\alpha}_{M}=\vec{\alpha}_{R}  - b\vec{h}_i$};
\draw (3.3,0)--node [] {} (5,0);
\draw (5,0) -- (6,1.2) node [right] {$\vec{\alpha}_{2211}=-b \vec{\rho}$};
\draw (5,0) -- (6,-1.2) node [right] {$\vec{\alpha}_L = \vec{\alpha}_{R}-b \vec{\omega}_1$};
\end{tikzpicture}
\end{equation}
We are interested here in the structure constants
\begin{equation}
C_{-b \vec{\rho}, \vec{\alpha}_M}^{\vec{\alpha}_L} = 
\left<V_{2\vec{\alpha}_0-\vec{\alpha}_{L}}(\infty)V_{-b \vec{\rho}}(1) V_{\vec{\alpha}_{M}}(0)\right>.
\end{equation}
From the neutrality condition \eqref{neutrality} and using formula \eqref{dotint}, one obtains,

\begin{eqnarray}
C_{-b \vec{\rho}, \vec{\alpha}_L+b \vec{\rho}}^{\vec{\alpha}_L} &=& 1,
\nonumber \\
C_{-b \vec{\rho}, \vec{\alpha}_L+b \vec{e}_1}^{\vec{\alpha}_L} &=& \pi \; 
\frac{\gamma(1+b^2)\gamma(1+y+ b^2)}{\gamma(2+y+2b^2)} 
\label{2node}
\end{eqnarray}

\noindent The  results coincides with  Eq 1.56 of \cite{fl07c} 
\footnote{\, 
In Eq 1.56 of \cite{fl07c} we point out a typos: the second 
product on the r.h.s runs from $i+1$ to $n$ and not to $n-1$}. 
More complicated is the case when $\alpha_M = \alpha_L$, related to the presence of multiplicities.
The neutrality condition \eqref{neutrality} is satisfied with one screening of type $e_1$ and one 
screening of type $e_2$. One has

\begin{equation}
C_{-b \vec{\rho}, \vec{\alpha}_L}^{\vec{\alpha}_L} = 
\int \ll d^2 \;t_1 \rr 
\int \ll d^2 \;t_2 \rr
|t_1|^{2 b^2}|t_2|^{2 b^2}|t_1-1|^{2 x +2 b^2}|t_2-1|^{2 y}|t_1-t_2|^{2 b^2}
\end{equation}

\noindent In the case $y = x+b^2$, the above integral is computed in Equation \textbf{B.9} of  
\cite{dofa_npb85} and one obtains 

\begin{equation}
\label{intsimpl}
C_{-b \vec{\rho}, \vec{\alpha}_L}^{\vec{\alpha}_L} = 
2 \pi^2 \frac{\gamma(b^2)}{\gamma(b^2/2)}\prod_{i = 0, 1} 
\frac{\gamma(1 + (2+ i)b^2/2) \gamma(1 + y + i b^2/2)
}{ 
\gamma(2 + (3 + i)b^2/2+y)} \quad \text{for} \quad x+b^2=y
\end{equation}

\noindent In the most general situation, where $x+b^2\neq y$, the above integral 
can be computed using the procedure explained in the Appendix A of \cite{DoPiPu95}. 
The integral is then expressed as a quadratic combination of ${}_3 F_{2}$ hypergeometric 
functions computed at $z=1$. We define a vector $J=(J_1,J_2)$ where

\begin{equation}
J_1 = N_1 \;{}_3 F_2 \ll -y, 2 + 3 b^2 , 1 + b^2; 3 + 4 b^2 + x, 2 + 2 b^2| 1 \rr
\end{equation}  

\begin{equation}  
J_2 = N_2 \;{}_3 F_2 \ll -b^2, 2 + x+y+ 2b^2, 1 + x+b^2; 3 +3 b^2 + x+y, 2 + b^2+ y| 1 \rr,
\end{equation}

\noindent with

\begin{eqnarray}
N_1 & = & 
\frac{
\Gamma(2+3 b^2) \Gamma(1 + x+b^2) 
\Gamma(1 + b^2) \Gamma(1 + b^2)
}{
\Gamma(3 + 4 b^2+x) \Gamma(2 + 2b^2)
}, 
\nonumber 
\\
N_2 & = & 
\frac{
\Gamma(2+2b^2+x+y) \Gamma(1 +b^2) \Gamma(1 + y) \Gamma(1 + b^2)
}{
\Gamma(3 + 3 b^2+x+y) \Gamma(2 + b^2+y)
}  
\end{eqnarray}

Using the notation $[x]=\sin(\pi x)$, we introduce the $2\times 2$ matrices

\begin{equation}
M_1=
\ll
\begin{array}{cc}
\left[3b^2\right]  & \left[x+2b^2\right] \\
\left[y+b^2\right] & \left[y+2 b^2\right]
\end{array}
\rr,
\end{equation}
\begin{equation}
M_2=
\ll
\begin{array}{cc}
\frac{[b^2]^2}{[y+2b^2]} & \frac{[b^2][2b^2]}{[y+2b^2]}   \\
 \frac{[b^2][2b^2]}{[x+3b^2]} & \frac{[b^2]^2}{[x+3b^2]}
\end{array}
\rr\;,
\end{equation}
and $M = M_1^{-1}M_2$. We obtained the following expression for the structure constant,

\begin{multline}
\label{intdiff}
C_{-b \vec{\rho}, \vec{\alpha}_L}^{\vec{\alpha}_L} = 
\left[ b^2\right] \left[x+b^2\right] 
J_T M J 
\\
+ \sum_{j=1,2}
\ll 
\left[x+b^2\right]\left[y+b^2\right] J_{1} M_{2 j} J_j+ 
\left[x + 2b^2 \right]
\left[y        \right] J_2 M_{1 j} J_j 
\rr
\end{multline}

\noindent We have verified that the \eqref{intdiff} coincides with \eqref{intsimpl} when 
$x+b^2=y$.

\section{A $\cW_3$ basis}
\label{appendix.D.basis}

\noindent \textit{
We outline an algorithm to expand any $\cW_3$ state in terms of the basis states.
The same operations allow us to compute the matrix elements 
in Section \textbf{\ref{section.03.matrix.elements}}, following \cite{kms10}
} 

\subsection{Definitions}

We start with a number of simple definitions, all of which are self-evident, 
but we include them for completeness.

\noindent \textbf{Modes and products.}
We refer to 
$L_m$ as an $L$-mode, and to 
$W_n$ as an $W$-modes. 
The indices $m$ and $n \in \ZZ$ are mode-numbers.
We refer to 
a product of $L$-modes only as an $L$-product,
a product of $W$-modes only as a  $W$-product, and to
a product of $L$-modes and $W$-modes as an $LW$-product.
When an $LW$-product consists of 
one or more sequences of consecutive $L$-modes, 
followed by 
sequences of consecutive $W$-modes, \textit{etc.}, we refer to each of these 
sequences as an $L$-sub-product, and $W$-sub-product.

\smallskip
\noindent \textbf{Normal order.} 
An $L$-product $L_{m_1} \cdots L_{m_L}$ is normal-ordered if 

\begin{equation}
m_1 \leq m_2 \leq \cdots \leq m_L
\end{equation}

\noindent that is, the mode-numbers increase from left to right. 
Similarly, a $W$-product $W_{n_1} \cdots W_{n_W}$ is normal-ordered if 

\begin{equation}
n_1 \leq n_2 \leq \cdots \leq n_W
\end{equation}

\noindent An $LW$-product is normal-ordered if all $W$-modes act first 
from the left on the highest-weight state, in normal-order, then all 
$L$-modes act second, also in normal-order.  

\smallskip
\noindent \textbf{Inversion numbers.} 
An inversion number of an arbitrary product is the number of inversions,
or permutations of nearest-neighbouring modes, that are required to put 
a product in a specific form. To quantify the degree of disorder of 
an arbitrary $LW$-product, we use four inversion numbers, 

\smallskip
\noindent \textbf{1.} 
$I_{LW+}$ is the inversion number of non-negative $L$-modes with respect 
to the $W$-modes. 
For each non-negative $L$-mode in an $LW$-product, we record the number 
of $W$-modes on its right. The sum of these numbers is $I_{LW+}$ 

\smallskip
\noindent \textbf{2.} 
$I_{LW-}$ is the inversion number of     negative $L$-modes with respect 
to the $W$-modes. 
For each     negative $L$-mode in an $LW$-product, we record the number 
of $W$-modes on its left.  The sum of these numbers is $I_{LW-}$

\smallskip
\noindent \textbf{3.} 
$I_{WW}$ is the inversion number of $W$-modes with respect to each other. 
For each              $W$-mode in an $LW$-product, we record the number 
of lower $W$-modes on its right. The sum of these numbers is $I_{WW}$

\smallskip
\noindent \textbf{4.} 
$I_{LL}$ is the inversion number of $L$-modes with respect to each other. 
For each              $L$-mode in an $LW$-product, we record the number 
of lower $L$-modes on its right. The sum of these numbers is $I_{LL}$

\smallskip
\noindent \textbf{Basis states.}
The Hilbert space of a $\cW_3$ conformal field theory is spanned by a basis, 
the elements of which are created by the action of the negative $W$-modes in 
normal-order, followed by the action of the negative $L$-modes, also in 
normal-order,

\begin{multline}
| Y_L, Y_W, h, w \rangle = 
	L_{m_1} \cdots L_{m_L} 
	W_{n_1} \cdots W_{n_W} | h, w \rangle, 
	\\
m_1 \leq \cdots \leq m_L < 0, 
\quad
n_1 \leq \cdots \leq n_W < 0
	\label{basis.state}
\end{multline}

\noindent where 
$m_i < 0$, $i=1, \cdots, L$, 
$n_j < 0$, $j=1, \cdots, W$, 
$Y_L$ is a Young diagram with parts 
$| m_L | \leq \cdots \leq | m_1 |$, and
$Y_W$ is a Young diagram with parts 
$| n_W | \leq \cdots \leq | n_1 |$.
The quantum numbers $h$ and $w$ label the highest-weight state of the $\cW_3$ 
highest-weight representation that the state $ |Y_L, Y_W, h, w \rangle$ belongs 
to.

\smallskip
\noindent \textbf{Disordered states.} Any state that is \textit{not} in the 
form (\ref{basis.state}) is a disordered state. We encounter disordered states 
in intermediate steps of computations, including those of matrix elements 
of descendant states, as in Section \textbf{\ref{section.03.matrix.elements}}.
The $L$-modes and $W$-modes in disordered states can be non-negative.

\subsection{Expanding a disordered state in terms of basis states}
Following \cite{fl88}, any disordered state can be expressed as a linear combination 
of basis states. In Virasoro conformal field theories, the commutation relations 
are relatively simple, and can be used in a straightforward way to expand any
disordered state in terms of the basis states. 
In $\cW_N$ conformal field theories, the commutation relations are more involved, 
which is related to the fact that $\cW_N$, $N = 3, 4, \cdots$, is not a Lie algebra. 
It is instructive to see how one can expand any disordered state in terms of
basis states, in the case of $\cW_3$ theories. 
We outline one way to do this, based on a systematic application of the $\cW_3$ 
commutation relations in Section \textbf{\ref{w3.algebra}}. 
For the purposes of the algorithm that we outline in this appendix, it is 
convenient to re-write the commutation relations in a simple form that discards 
the coefficients.

\subsubsection*{$[L, W]$, the $L$-mode is non-negative.}
\label{LW}

\begin{equation}
L_m W_n \sim W_n L_m + W_{m+n}, \quad 0 \leq m, 
\label{LW.simple.commutator}
\end{equation}
\noindent that is, commuting a non-negative $L$-mode, from the left to the right 
of $W_n$, for any $n \in \ZZ$, we end up with two terms.
We use this commutation to move a non-negative $L$-mode, from the left to the right 
of a $W$-mode. 
The first  term has a lower $I_{LW+}$ inversion number than the term on the left. 
The second term can be interpreted as the absorption of $L_m$ into $W_n$ to produce 
$W_{m+n}$, which leads to a shorter product of modes, the normal-ordering of which
is a simpler problem than the one that we started with, thus the second term is 
also an improvement on the term on the left.

\subsubsection*{$[W, L]$, the $L$-mode is negative.}
\label{WL}

\begin{equation}
W_n L_m \sim L_m W_n + W_{m+n}, \quad m < 0,
\label{WL.simple.commutator}
\end{equation}
\noindent that is, commuting a negative $L$-mode to the left of $W_n$, $n \in \ZZ$, 
we end up with two terms.
Each term on the right is an improvement on the term on the left for the same
reasons as in section \textbf{\ref{WL}}. 

\subsubsection*{The commutator of two $W$-modes.}
\label{WW}

\begin{equation}
W_n W_m \sim 
W_m W_n + 
L_{m+n} + \delta_{m+n, 0} + 
\sum_{k=-\infty}^\infty : L_{-k} L_{m+n+k} :,
\quad 
m < n
\label{WW.simple.commutator}
\end{equation}

\noindent that is, commuting a higher-mode $W_n$ to the right of a lower-mode 
$W_m$, we end up with four terms. Each of the first three terms on the right 
is an improvement on the term on the left for the same reason as in
section \textbf{\ref{WL}}. The fourth term is an improvement in the sense that it 
replaces two $W$-modes by two $L$-modes. The resulting state is easier to
normal-order than the one that we started with, since $L$-modes obey simpler
commutation relations than the $W$-modes. In fact, the current $L$-modes are
normal-ordered with respect to each other.

Following an application of
(\ref{WW.simple.commutator}), the initial state is replaced by a sum 
over infinitely-many states. To be able to expand a disordered state in terms of 
a linear combination of finitely-many basis states, in finitely-many number of
steps, we must make sure that the sum, produced by (\ref{WW.simple.commutator}) 
is finite. While the number of states produced in (\ref{WW.simple.commutator}) 
is formally infinite, only finitely-many products survives. This is because, 
except in finitely-many cases, all new states contain a mode that is
sufficiently-positive to kill the state it is in. 
This is because the number of $W$-modes to the right of the new pair of $L$-modes 
is finite, and the degree of each of these modes, whether positive or negative, 
is also finite. 
Using (\ref{LW.simple.commutator}), a sufficiently-positive $L$-mode can be moved 
to the right, producing either a sufficiently-positive $L$-mode or 
a sufficiently-positive $W$-mode that kills the highest-weight state.

While the number of states increases, each descendant state is an improvement 
over its ancestor. Because the initial degree of disorder is finite, the 
increment in the number of states is finite, and the disorder decreases by 
a finite amount at each step, the algorithm will terminate in finitely-many 
steps. 

\subsubsection*{The commutator of two $L$ modes.}
\label{LL}

\begin{equation}
L_n L_m \sim L_m L_n + L_{m+n} + \delta_{m+n, 0},
\quad
m < n
\label{LL.simple.commutator}
\end{equation}

\noindent that is, commuting the higher-mode $L_n$ to the right of the lower-mode 
$L_m$, we end up with two terms. Each of these terms is an improvement on the term 
on the left for the same reason as in section \textbf{\ref{LW}}.

\noindent \textbf{Finitely-many steps.} 
\label{finiteness}
We consider finite-level $\cW_3$ states, constructed by the action of finitely-many
$L$-modes and finitely-many $W$-modes, all with finite mode-numbers. We emphasise
finiteness because the algorithm that we propose is iterative. For an iterative
algorithm to make sense, it must terminate after a finitely-many steps. 
To show that the algorithm that we propose is finite, we use the inversion 
numbers, defined above, to measure how far we are from our goal. In this appendix, 
our goal is to expand an arbitrary disordered $\cW_3$ state in terms of 
basis $\cW_3$ states.

\noindent \textbf{The algorithm.}
We outline an iterative algorithm. Each iteration takes an $LW$-product with 
a finite degree of disorder as an input, and generates finitely-many descendant
$LW$-products as outputs.
Each output $LW$-string has 
\textbf{1.} a lower degree of disorder, 
\textbf{2.} a smaller number of $L$-modes, or
\textbf{3.} a smaller number of $W$-modes, 
than the input $LW$-product. 
After a finite number of steps, every output $LW$-product is normal-ordered, 
and the algorithm terminates. 

\noindent \textbf{Four steps.} The algorithm is based on the iteration of four 
operations, based on the four simplified commutators
(\ref{LW.simple.commutator}), 
(\ref{WL.simple.commutator}), 
(\ref{WW.simple.commutator}), and 
(\ref{LL.simple.commutator}). 

\smallskip
\noindent \textbf{Step 1.} 
We scan the input $LW$-product from right to left, and locate the first 
non-negative $L$-mode, which has a negative $L$-mode or any $W$-mode to 
its right. 
We use (\ref{LW.simple.commutator}) to move this mode one step to the right. 
If the mode to the right is a negative $L$-mode, we use (\ref{LL.simple.commutator}) to move 
the non-negative $L$-mode to the right again. 
Repeating step \textbf{1} finitely-many times, the result is a linear 
combination of $LW$-products that has no non-negative $L$-modes.

\smallskip
\noindent \textbf{Step 2.} 
We scan the state from left to right, we locate the first negative 
$L$-mode, which has any $W$-mode to its left. 
We use (\ref{WL.simple.commutator}) to move this mode one step to the left. 
Repeating step \textbf{2} finitely-many times, the result is a linear 
combination of $LW$-products. Each of these products consists of 
a generally disordered $L$-product, with negative $L$-modes only, 
on the left, and a generally disordered $W$-product on the right.

\smallskip
\noindent \textbf{Step 3.} Consider each of the $LW$-products obtained 
at the end of the final iteration of step \textbf{2} above. 
Scan the $LW$-product from right to left, and 
locate the first positive $W$-mode, which has a negative $W$-mode to 
its right. 
Use (\ref{WW.simple.commutator}) to move this mode one step to the right. 

\smallskip
\noindent \textbf{Repeating steps 1 and 2.} 
One of the descendants produced in step \textbf{3} is a two-operator 
$L$-product in the middle of a previously-pure $W$-product. To clear 
this, we need to repeat step \textbf{1} and step \textbf{2} again. 
Iterating steps \textbf{1} and \textbf{2} finitely-many times, we end 
up with $LW$-products, such that each of these products consists of 
a disordered $L$-product to the left, and a normal-ordered $W$-product 
to the right. 

\smallskip
\noindent \textbf{Step 4.} 
We consider each $LW$-product produced in the final iteration of step 
\textbf{3}, and use (\ref{LL.simple.commutator}) to order the $L$-sub-product. We end up 
with a set of $LW$-products such that the $L$-product is normal-ordered 
and the $W$-product is normal-ordered. 
This concludes the algorithm.

\section*{Acknowledgements}
We thank the Institut Henri Poincare, Paris, where this work was completed, 
for excellent hospitality and financial support. 
The work of V.B. was performed at the Landau Institute for Theoretical Physics,
with financial support from the Russian Science Foundation (Grant No.14-12-01383). 
O. F. is supported by the Australian Research Council. 
We thank T. Dupic, J. Gomis, Y. Ikhlef, Y. Matsuo, S. Ribault and A. Tanzini for 
discussions, P. Boalch and Y. Haraoka for explanations of important aspects of 
Fuchsian systems, X. Cao for preliminary contributions to the study of null-vectors 
of fields in the adjoint representation, and B. LeFloch for discussions and 
comments on an earlier version of the manuscript. 

\bibliographystyle{JHEP}
\bibliography{bibnonwy.bib}

\end{document}